%% file: review.tex
\documentclass[a4paper,12pt]{iopart}

\usepackage{hyperref}
\hypersetup{
  colorlinks   = true, 
  urlcolor     = blue, 
  linkcolor    = blue, 
  citecolor   = red 
}
\usepackage{iopams} 	
\usepackage{setstack} 	
\usepackage{bbm}
\usepackage{graphicx}	
\usepackage{subfig}
\usepackage{wasysym}

\usepackage[normalem]{ulem}   
\usepackage[usenames]{color}  

\eqnobysec				

\input{shortcuts}

\begin{document}

\review{Distribution of entanglement in large-scale quantum networks}

\author{S Perseguers, G J Lapeyre Jr$^1$, D Cavalcanti$^{1,2}$,
M Lewenstein$^{1,3}$, and A Ac\'in$^{1,3}$}

\address{$^1$ ICFO--Institut de Ci\`encies Fot\`oniques, Mediterranean
Technology Park, 08860 Castelldefels, Spain}

\address{$^2$ Centre for Quantum Technologies, National University of Singapore,
Block S15, 117543 Singapore, Singapore}

\address{$^3$ ICREA-Instituci\'o Catalana de Recerca i Estudis Avan\c cats,
Lluis Companys 23, 08010 Barcelona, Spain}


\begin{abstract}
 The concentration and distribution of quantum entanglement
 is an essential ingredient in emerging quantum information
 technologies. Much theoretical and experimental effort has
 been expended in understanding how to distribute
 entanglement in one-dimensional networks. However, as
 experimental techniques in quantum communication develop,
 protocols for multi-dimensional systems become
 essential. Here, we focus on recent theoretical
 developments in protocols for distributing entanglement
 in regular and complex networks, with particular
 attention to percolation theory and network-based error
 correction.
\end{abstract}

\pacs{03.67.Bg, 03.67.Hk, 03.67.Pp, 03.67.Mn, 64.60.ah}

\setlength{\textheight}{10.0in}	
\maketitle
\tableofcontents
\markboth{\sc Contents}{}		
\setlength{\textheight}{9.3in}	

\graphicspath{{figures/}}
\input{sections/introduction}
\input{sections/concepts}
\input{sections/percolation}
\input{sections/correction}
\input{sections/complex}
\input{sections/conclusion}

\ack \markboth{\sc Acknowledgments}{} 
This work was supported in part by the Spanish MICINN
(TOQATA, FIS2008-00784), by the ERC (QUAGATUA), and EU
projects AQUTE and NAMEQUAM. 
AA is supported by an ERC Starting Grant PERCENT, the EU
project Q-Essence and the Spanish Project FIS2010-14830.
DC acknowledges financial support from the
National Research Foundation and the Ministry of Education
of Singapore and thanks A. Grudka and M. Horodecki for
correspondence. GL thanks A. Fowler for correspondence.
\section*{References}
\markboth{\sc References}{}
\bibliographystyle{unsrt} 
\bibliography{review}

\end{document}

%% file: shortcuts.tex
\newcommand{\ie}{\textit{i.e.}}

\newcommand{\eps}{\varepsilon}
\newcommand{\id}{\rm{id}}

\newcommand{\cluster}{\mathcal{C}}
\newcommand{\rhoW}{\rho_{\rm{W}}}

\newcommand{\prob}{{\rm prob}}

\newcommand{\GHZ}{{\rm GHZ}}

\newcommand{\cB}{\mathcal{B}}

\newcommand{\cL}{\mathcal{L}}

\newcommand{\cN}{\mathcal{N}}

\renewcommand{\vec}{\boldsymbol}

\DeclareSymbolFont{symbolsC}{U}{txsyc}{m}{n}
\SetSymbolFont{symbolsC}{bold}{U}{txsyc}{bx}{n}
\DeclareFontSubstitution{U}{txsyc}{m}{n}
\DeclareMathSymbol{\Connected}{\mathrel}{symbolsC}{19} 
\newcommand{\connected}{\!\Connected\!}

\newcommand{\bra}[1]{\langle #1|\,}
\newcommand{\ket}[1]{\,|#1 \rangle}

\newcommand{\ketbra}[2]{\,|#1 \rangle\!\langle #2|\,}

\newcommand{\Ket}[1]{\,\big{|}#1 \big{\rangle}}

%% file: sections/introduction.tex
\newpage
\setcounter{footnote}{0}
\section{Introduction}
\markboth{\sc Introduction}{}

The idea of quantum entanglement has a long history
\cite{EPR35,SCH35a,SCH35b,Bel64}, although an intensive
search for a comprehensive theory of entanglement only arose
with quantum information theory. This search grew out of the
realisation that quantum entanglement is an essential
resource for developing information technologies that are
radically different than those possible in a purely
classical world. In fact, when two physical systems are
sufficiently entangled, they exhibit correlations that are
stronger than possible with any classical theory. These
strong correlations can then be exploited by cryptographic,
communication, or computation protocols. As with classical
information theory, there is a fundamental need to
understand how to distribute information, that is, how
to transmit a signal between two parties.  But quantum
mechanics imposes severe limitations, both fundamental and
practical, on copying, encoding, and reading
information. Thus, distributing quantum entanglement is an
extremely challenging problem. This review presents work
that attempts to meet that challenge.

Knowledge of the network in which this entanglement distribution will
take place is still in a nascent stage. Naturally, most attention was
initially focused on one-dimensional
setups~\cite{BDCZ98,DBCZ99,CTSL05}.  However, it is natural to
consider distribution on multi-dimensional or otherwise more
highly-connected networks, whose structure may be designed explicitly
for distribution or be imposed by geographical constraints.  Obvious
examples of connections to existing bodies of work that will become
increasingly important are the classical theory of complex networks,
particularly those with an internet-like
structure~\cite{WS98,AB02,Wat04,NBW06}. From another point of view,
the production and manipulation of entanglement on micro- or
nano-scale networks is progressing rapidly~\cite{Kimble08}.  In this
case, we will likely be presented with regular arrays of elements that
can be entangled.  Both of these kinds of systems, and others not yet
imagined, will require an understanding of entanglement distribution.

In any communication application, information is encoded in the state
of a physical system.  As this system travels from the sender to the
receiver, it interacts with the environment, and a degradation or
eventually a loss of information may occur. In classical systems,
devices such as amplifiers have been specially developed to overcome
this problem, by repeatedly copying the information content that is
being transmitted.  However, when single atoms or photons are used as
information carriers, one faces a fundamental property of quantum
mechanics which makes quantum communication challenging: quantum
information cannot be copied perfectly \cite{WZ82}.  On the other
hand, quantum entanglement opens new possibilities in manipulation of
information that are fundamentally impossible in the classical
theory \cite{NC00}.

Before giving a detailed description of entanglement, we
briefly review a few of the most important applications of
this phenomenon; we refer the reader to~\cite{HHHH09} for a more complete review.
These examples will repeatedly refer to
the fundamental quantum system for quantum information
science, the {\em qubit}, which is used as a quantum analogue of the
classical binary digit or {\em bit}. The qubit is an
abstraction that is realised by any two-level
quantum system: the spin of an electron or neutron, the
polarisation of a photon, or the first two energy levels of an atomic electron
in a resonant field, just to name a few. These systems can be considered to have a single,
two-dimensional degree of freedom, which means that,
for a given orientation of the measurement device, a measurement
always gives one of two results.

\paragraph{Quantum teleportation} It has been proven that
the unitary evolution of quantum mechanics implies that an
unknown quantum state cannot be duplicated or
cloned~\cite{WZ82}. However, an unknown quantum state can be
transported over an arbitrarily long distance as long as an
auxiliary perfectly entangled pair of particles (also called
a Bell pair) and a classical communication channel are
established over the same distance~\cite{BBC+93}. The
entangled pair is created via a local temporary interaction
between two qubits, which are then separated by the desired
distance.
The procedure is as follows. Two distant parties,
traditionally called Alice and Bob, share an entangled pair
of qubits, while Alice has an additional ``data'' qubit that
she wishes to send to Bob. Alice performs a certain
measurement on the two qubits she possesses and communicates the result
classically to Bob. Bob then applies a transformation on his qubit in a manner
prescribed by the message from Alice. The result is that
Bob's qubit is now in the original state of the data qubit
of Alice, which meanwhile has lost its information content.

\paragraph{Quantum distributed computing}
 Distributed computing consists of several nodes
that do computations independently while periodically
sharing results. Entanglement appears in several places in
quantum distributed computing protocols, including in the
input states or in the communication channels used in
sharing results between nodes. Attempts have been made to
design quantum distributed computers so that limited
entanglement resources are spread in an optimal way among
components~\cite{CEHM99}.

\paragraph{Quantum key distribution} Classical public key
cryptography schemes are widely used, for instance, in internet
security algorithms.  Entangled pairs can be used to securely generate
and distribute the classical private key necessary for these
schemes~\cite{Eke91,SBCD09} (although performing quantum key
distribution without distributing entanglement is also
possible \cite{BB84}). Using a quantum protocol, Alice and Bob
generate a series of random bits, sharing knowledge of the results,
but preventing others from eavesdropping. To this end, they initially
share a number of Bell pairs, which they measure sequentially, each
time choosing an orientation randomly from a pre-determined set. They
can use a portion of their results to compute a bound on the amount of
eavesdropping (or noise) that has taken place, and another portion to
generate the key.

\paragraph{Superdense coding} If two parties share a Bell
pair, then \textit{two} classical bits of information can be
sent from one party to the other one, even though each party
physically possesses only \textit{one} qubit~\cite{BW92}.
To accomplish this, Alice applies one of four previously
agreed-upon unitary operations to her qubit. A unitary
operation, in contrast to measurement, transforms the qubit
in a non-destructive and coherent way. This transforms the
Bell state to one of four orthogonal states that together
form the Bell basis.  Alice sends her qubit to Bob, who then
performs a joint measurement on both qubits, thus
distinguishing reliably among the four messages Alice can
send.

\paragraph{}
Each of the applications mentioned above requires entangled pairs of
particles to be generated and distributed between two distant parties.
Currently, the main technological difficulty is to create remote
entanglement which, in most experiments, is achieved by sending
polarised photons through optical fibres. In fact, due to noise,
scattering, and absorption, the probability that the quantum
information contained in such photons reaches its destination
decreases exponentially with the distance. Another experimental
challenge is to transfer the quantum state of the photon onto that of
a quantum memory, such that the entanglement can be manipulated and
stored; see~\cite{SAA+10} for a recent review on quantum memories.

\paragraph{Outline}
We first review some well-known results on entanglement and
describe the operations that allow one to propagate quantum
information in a network, such as entanglement 
purification and
entanglement swapping; see \sref{concepts_entanglement} and
\sref{concepts_QN} \cite{HHHH09}. Based on these operations, the quantum
repeaters enable entanglement to be generated over a large distance in
one-dimensional networks \cite{BDCZ98}; see \sref{concepts_repeaters}.
However, in order to obtain reasonable communication rates, they
require an amount of entanglement per link of the network that
increases with the distance over which one would like to distribute
entanglement, which is out of reach with present technologies.
A natural question is whether the higher
connectivity of the stations or nodes of more complex networks can provide
some advantage over a one-dimensional setup when distributing
entanglement; see \sref{percolation}. The first protocol exploiting
this fact uses ideas of percolation in two-dimensional lattices:
for pure states, if enough entanglement is generated between
neighbouring stations, then it can be propagated over infinite
distance \cite{ACL07}; see \sref{percolation_pure}. 
For general mixed states, \ie\ quantum states that contain random noise,
this result no longer holds.
For some specific types of noise, however,
entanglement percolation still
allows entanglement to be generated between infinitely distant
stations~\cite{BDJ09,BDJ10}; see \sref{percolation_mixed}.

It turns out that all percolation strategies need, in the end, perfect
operations to be applied on the system.  The study of strategies with
perfect operations is certainly useful, for instance in establishing
fundamental limits. Still, we must ask if the higher connectivity of
quantum networks will be of practical use, given that, in realistic
scenarios, operations are not perfect, but rather introduce noise.  We
will see that the answer to this question is positive.  But, in order
to accommodate noisy operations, radically different protocols, based
on error correction, must be
designed~\cite{PJS+08,FWH+10,Per10a,LCK12,GHH+12};
see \sref{correction}. In fact, while entanglement percolation relies
on the existence of \textit{one} path of perfectly entangled states
between any two stations, network-based error correction extracts its
information from
\textit{all} paths connecting the two stations. Contrary to the
quantum repeaters, no quantum memory is needed in that case.

Initially applied to regular lattices, the study of
entanglement distribution has since been extended to complex
networks~\cite{PLAC10}; see \sref{complex}. This is a natural generalisation
because present communication networks exhibit a complex
structure.  Furthermore, while mostly focused on pure
states, recent work concerns the manipulation of
entanglement in noisy quantum complex networks;
see \sref{complex_complex_mixed}. These results emphasise the
fact that the quality of the quantum communication between
two stations will depend greatly on our understanding of the
interplay of the network topology and the quantum operations
available on the system.


%% file: sections/concepts.tex
\newpage
\setcounter{footnote}{0}
\section{Concepts and methods}
\markboth{\sc Concepts and methods}{}
\label{concepts}

In this section we describe some well-known results on entanglement and the
basic operations that are used to propagate quantum information in a network;
see~\cite{P95} for a thorough treatment of this material in a pedagogical
setting. A reader familiar with quantum information theory may skip
this part and jump to the discussion of quantum repeaters
in \sref{concepts_repeaters} or percolation in \sref{percolation}.

\subsection{Quantum states}
\label{concepts_measurement}

The state of a single qubit can be written as
\begin{equation}\label{one_qubit}
\ket{\phi}= \sqrt{\alpha_0}\ket{0} + e^{i \theta}\sqrt{\alpha_1}\ket{1},
\end{equation}
where $\ket{0}$ and $\ket{1}$ represent a choice of {\em basis},
called the {\em computational} basis, in a Hilbert space. Since the
phase $\theta$ is irrelevant to the remainder of this review, we
shall choose $\theta=0$.  We say that the system is in a {\em
coherent superposition} of the two basis states.  We can measure the
state of the qubit in this basis, which corresponds, for instance, to
the orientation of magnets acting on the spin of electrons in the
laboratory. The probability that we measure $0$ is $\alpha_0$ and
the probability that we measure $1$ is $\alpha_1$.  We must have
$\alpha_0+\alpha_1=1$, as these are the only possible outcomes. Upon
measurement in the computational basis, the state of the system
collapses into one of the two basis states, say $\ket{0}$.  If we now
repeat the same measurement, we have $\alpha_0=1$ and $\alpha_1=0$, so
that we obtain $\ket{0}$ with probability $1$. However, if we rotate
our magnets to an orientation corresponding to a different basis and
then measure, we must expand $\ket{0}$ in that new basis in order to
calculate the probabilities of obtaining each of the two possible
results.

\subsubsection{Measurements and quantum evolution}%
While the study of quantum measurement is a broad and deep
subject, here we only need to introduce a few ideas to discuss
entanglement distribution. Measurements occur in the laboratory,
but the computational tool to predict their result associates with
each type of measurement a set of linear operators on the state
space. The measurements described above are called {\em
projective} measurements, which are defined by a collection of
projectors $E_j$  onto subspaces of the state space, each of which
is associated with a possible measurement value. After measuring a
value, we know that the system has collapsed into a state in the
subspace corresponding to the associated projector. The projectors
are orthogonal, and the requirement that we must get some result
implies that their sum is the identity. The maximum number of
orthogonal projectors is equal to the dimension of the state
space and corresponds to a complete measurement.
On the contrary, we use an {\em incomplete} measurement, with
fewer projectors, when we want to extract only partial information from a state
while preserving some property common to all subspaces. Finally,
we will sometimes make use of {\em generalised} measurements, in
which the condition that the operators are orthogonal is relaxed.
Measurements are then defined by a collection of
semi-definite positive operators that sum to the identity~\cite{NC00}.

The other component of quantum theory that transforms states is evolution.
Evolution is also represented theoretically by operators, but in contrast
with measurement, an initial state is transformed into a final state in
a deterministic way. Operators representing evolution must have the
algebraic property of unitarity in order to conserve
total probability.
These operators are commonly referred to as {\em unitaries}. 
The entanglement distribution procedures
in this review are built from a combination of measurements and
unitaries on quantum states, together with classical communication.

\subsubsection{Entanglement}
\label{concepts_entanglement}

To illustrate the basic idea of entanglement while making a connection
with the remainder of this review, let us consider two qubits, \ie\
two systems each of which consist of a two-level quantum state. Two
qubits live in a four-dimensional state space whose computational
basis is $\ket{00},\ket{01},\ket{10},\ket{11}$, where the first binary
digit corresponds to qubit $A$ and the second to qubit
$B$. Mathematically, these basis vectors are tensor products of
vectors in the local spaces: $\ket{ij} = \ket{i}\otimes\ket{j}.$ If the two systems have never
interacted directly or indirectly, then each one has an
independent description of its state, as in
\eref{one_qubit}. In this case, it is possible to find local bases for
each qubit such that the joint state is one of the four computational
basis states of the joint space. These states are called product
states. Measurements and operations applied on one system have no
effect on subsequent measurement outcomes on the other system.
However, suppose an interaction between the systems is turned on, then
off. In general, there is no local basis such that the state of the
system after the interaction can be written as a factor of states of
the subsystems.  A pure bipartite quantum state is then said to be
{\it entangled} if it cannot be written as a tensor product. That is,
it cannot be written as
\begin{equation}
    \ket{\phi_{AB}} = \ket{\phi_A} \otimes \ket{\phi_B}.
    \label{eq:def_ent_pure}
\end{equation}
A famous example of entangled states are the four Bell states
\numparts
\begin{eqnarray}
    \ket{\Phi^{\pm}} &\equiv \frac{1}{\sqrt{2}}\left(\ket{00}\pm\ket{11}\right),\\
    \ket{\Psi^{\pm}} &\equiv \frac{1}{\sqrt{2}}\left(\ket{01}\pm\ket{10}\right),
\end{eqnarray}
\endnumparts
which form a basis of two-qubit systems.
These states are of central importance as they are
{\em  maximally entangled} states.
The projective measurement corresponding to the four states
of the Bell basis is called a Bell measurement.

Just as two two-dimensional systems form a four-dimensional system, in
general, the space describing a collection of quantum
systems of dimensions $n_1,n_2,\ldots, n_N$ has dimension
$n_1\times n_2\times \ldots \times n_N$. However, generalising the case of two
qubits, we may partition any of these composite spaces into
two subspaces, and viewed this way, the system is called a
{\em bi-partite} system. The study of bi-partite
entanglement, that is entanglement between the two
subspaces, is far better understood than the more general
case of multi-partite entanglement, and we shall mostly be
concerned in this review with bi-partite systems.

Until now, we have discussed only what are known as {\em pure}
quantum states, in which the probability of
measuring a value is of local quantum mechanical origin.
However, the generic case is that the system is entangled
with other systems which we cannot measure.  It turns out
that to predict local measurements in this case, we can
assume that the local state is effectively a classical
distribution of several pure states, 
and can be written as
\begin{equation*}
\rho = \sum_i p_i \ket{\psi_i}\bra{\psi_i},
\end{equation*}
where $\{p_i\}$ is a probability distribution.
This classical ensemble of pure states is known as a {\em mixed} state.
It is important to realise, however, that
this decomposition is not unique.
 These states are described not
by vectors in the Hilbert space of pure states, but rather by linear
operators known as density
operators, which act on the same Hilbert space.  For instance
the density operator corresponding to a pure state is
given by the outer product of the pure state vector with itself, which is
denoted by $\ket{\phi}\bra{\phi}$. Rules of quantum
mechanics together with classical statistics imply that
density operators
must be non-negative and have trace one.
The set of operators on finite-dimensional Hilbert spaces may be
represented by a set of matrices that depends on a chosen basis. 
Since all systems considered in this review
are composed of locally finite-dimensional Hilbert spaces, we follow
the common practice of using the term {\em density matrix} even
if the choice of basis is not fixed.

The notion of entanglement can be easily generalised to mixed
states~\cite{Wer89}. In this case, a bi-partite system is
entangled if and only if its joint density operator $\rho_{AB}$
cannot be written 
\begin{equation}
    \rho_{AB} = \sum_i p_i\ \ket{\psi_i^A}\bra{\psi_i^A}\otimes \ket{\phi_i^B}\bra{\phi_i^B},
    \label{eq:def_ent_mixed}
\end{equation}
that is, it is not a mixture of product states.

\paragraph{Measures of entanglement}
\label{concepts_entanglement_measures}

If one qubit of a Bell pair is measured in the computational
$Z$ basis, that is, if it is projected onto the eigenstates
$\ket{0}$ and $\ket{1}$ of the Pauli $Z$ matrix, then the
measurement outcome is either 0 or 1 with equal
probabilities. If the second qubit is then measured in the
same basis, the outcome is completely determined by the
first result. But, if we instead were to measure each qubit in the same arbitrary
rotated basis, the same correlation between outcomes would
be seen. In fact, it can be shown that the Bell states
are the maximally-correlated two-qubit states;
 they are maximally entangled. In operational terms this means that
they can be used to teleport perfectly the state of exactly one qubit.
In practice, however, the entanglement between two qubits is never perfect, so
that \textit{partially} entangled states have to be considered. In what follows
we present some common measures of entanglement.

In the pure-state formalism, local bases may be found so that
any two-qubit state is written
\begin{equation}
    \ket{\varphi} = \sqrt{\varphi_0}\ket{00}+\sqrt{\varphi_1}\ket{11},
    \label{eq:phi}
\end{equation}
where the two \textit{Schmidt coefficients} $\varphi_0$ and
$\varphi_1$ satisfy $\varphi_0+\varphi_1 = 1$ and
$\varphi_0\geq\varphi_1$ (by convention).  If one of the coefficients
$\varphi_0$ or $\varphi_1$ vanishes, the system is
separable. If $ \varphi_0 = \varphi_1 =1/2$, it is maximally
entangled. Otherwise, the system is said to be partially or
weakly entangled.
An important measure of entanglement is
\begin{equation}
    E(\varphi) = 2 \varphi_1 \in [0,1],
    \label{eq:SCP}
\end{equation}
which corresponds to the optimum probability of successfully converting
$\ket{\varphi}$ into a perfect Bell pair by local operations~\cite{Vid99}; see
\sref{concepts_entanglement_purification}.

Another common measure of entanglement, the
\textit{concurrence}~\cite{Woo98}, reads in the case of pure states:
\begin{equation}
    C(\varphi) = 2 \sqrt{\varphi_0\varphi_1} \in [0,1].
    \label{eq:CPure}
\end{equation}
For mixed state one can generalise the concurrence in a similar way as done for the entropy of entanglement. 
The mixed state of two qubits is represented by a
four-by-four density matrix, which requires, in general,
fifteen parameters. Any density matrix of two qubits $\sigma$ can
be transformed by local random operations to the standard form
\begin{equation}
    \rhoW(x) = x \ketbra{\Phi^+}{\Phi^+} + \frac{1-x}{4}\ \id_4,
    \label{eq:Wx}
\end{equation}
where $\id_4$ is the corresponding identity matrix.
This process is known as depolarisation, and $\rhoW$
is called a Werner state. It is important to notice that the depolarisation process typically increases the entropy of the system and reduces its entanglement. One sometimes
explicitly expands this expression in the Bell basis $\cB$:
\begin{equation}
    \rhoW(F) = \Big(F,\frac{1-F}{3},\frac{1-F}{3},\frac{1-F}{3}\Big)_{\cB},
    \label{eq:WF}
\end{equation}
where the components correspond to the weight of each Bell state in the mixed state decomposition of $\rhoW(F)$.
Although depolarisation may decrease entanglement,
it is done in a way that preserves the  {\it fidelity} with
respect to a preferred state~\cite{BBP+96}:
\begin{equation}
    F(\rhoW) \equiv \bra{\Phi^+}\rhoW\ket{\Phi^+}=(3x+1)/4.
\end{equation}
This fact is important because
the concurrence of the Werner state $\rhoW$ is related to the
fidelity via
\begin{equation}
    C(\rhoW) = \max\{0,2\,F(\rhoW)-1\}.
         \label{eq:Cmixed}
\end{equation}
The state $\rhoW$ is entangled if and only if its concurrence is
strictly positive, that is if $x>1/3$; or equivalently, if
\begin{equation}
    F>F_{\min} = 1/2.
    \label{eq:Fmin}
\end{equation}
Note that \eref{eq:Wx} emphasises the fact that a Werner state is
a mixture of a perfect quantum connection and a completely unknown
state. This observation is important, for
instance, when discussing the limitations of entanglement
propagation in noisy networks; see \sref{percolation_open}. The
Werner state is an example of a \textit{full-rank} state, that is,
one with no vanishing eigenvalues.
Full-rank states result from a general noise model and are thus
important because they model any state found in a laboratory. However,
the entanglement contained in these states cannot be extracted easily,
and special techniques have to be developed to this aim;
see~\sref{concepts_entanglement_purification}.

The discussion to this point may leave the reader with the impression
that characterisation of entanglement is a relatively uncomplicated
task. In fact it is a difficult and deep question, especially for
mixed and multipartite states~\cite{PV07}. On one hand a separable state can be written as a
classical mixture of projectors onto product states.  On the other
hand, there is no general algorithm to determine whether a given
density operator can be written in this form,
although significant progress has been made~\cite{DPS04, HHHH09}.

\subsection{Entanglement manipulation in quantum networks}
\label{concepts_QN}

In the introduction, we have seen that entanglement is a valuable resource for a
variety of quantum information applications. It is thus essential
to understand the non-trivial task of creating and
distributing entanglement between distant parties.  The
subject is greatly complicated, in fact largely determined,
by the fact that entanglement is a very fragile resource, in
the sense that it inevitably deteriorates while being
manipulated or stored. We shall see that the attempt to
overcome this difficultly has led naturally to the
consideration of network theory.

The most direct way to produce entanglement between spatially
separated parties is to entangle two particles locally and then to
send one of them physically to another location. Most research in
this direction involves sending photons through optical fibres,
which suffer from inherent limitations due to photon loss via
absorption as well as coherence of the state.
The limit at this time is about 100~km~\cite{Hubel07},
because the probability of
transmission decays exponentially with the distance,
becoming as low as $10^{-20}$ for
$1000$~km~\cite{SSRG11}.
Following the techniques developed
in classical systems in order to overcome similar but much
less severe limitations, creating entanglement between
distant stations via a series of intermediate nodes has been
proposed. The main and crucial difference with classical
information is that qubits cannot be copied, so that the
intermediate links have to be joined together in a very
subtle way, known as entanglement
swapping, which will be discussed below.
This one-dimensional system of links
and nodes has then a natural generalisation to a network of
arbitrary geometry.

\begin{figure}
    \begin{center}
        \includegraphics[width=3.2cm]{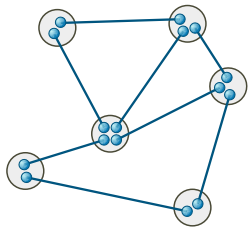}
        \caption{A quantum network. The stations (nodes)
            initially share some partially entangled pairs
            of qubits (links); for simplicity, we
            sometimes omit explicitly drawing the qubits at
            the stations. By LOCC, we mean that any
            quantum operation can be applied on the qubits
            of a station, but that only classical information
            is exchanged between the nodes.}
    \label{fig:qnetwork}
    \end{center}
\end{figure}

\subsubsection{Structure of quantum networks}
\label{structure_quantum_networks}

The quantum networks that we consider
consist of two basic elements: \textit{nodes} (or \textit{stations}), 
 each of which possesses one or more qubits; and
\textit{links}, each of which represents entanglement 
 between qubits on different nodes. This generalises
the one-dimensional scenario in that links may exist between
all pairs of nodes, rather than only neighbouring nodes on
a chain. An example of such a quantum network is shown in
\fref{fig:qnetwork}.
It is often useful to interpret this structure in
terms of graph theory, where the nodes become vertices and
the links become edges. Furthermore, we sometimes use the language
of statistical physics by referring to a regular graph with local
connections as a \textit{lattice}.

\subsubsection{Local operations and classical communication}
In manipulating entanglement in a quantum system, we typically begin with
a given distribution of entangled pairs of qubits and then apply a
series of operations designed to distribute the entanglement in a
useful way. The problems we consider naturally impose a
distinction between local and distributed resources.  It is
important to have a clear notion of what kind of operations are
allowed on these resources. In this review, we shall only consider local
operations and classical communication (LOCC) on the network.
It turns out that it is possible to provide an operational definition
of entanglement as the quantum resource that does not increase
under LOCC. It is easy to see that this definition is equivalent
to the pure mathematical one given by~\eref{eq:def_ent_mixed}.

It is now clear how the concept of LOCC leads to the quantum
network shown in \fref{fig:qnetwork}.
Recall that the network initially is composed of some entangled pairs of qubits,
with one party of each pair occupying a node of the network. However, several
qubits from different pairs occupy a single node along with
other possible resources, both quantum and classical. When we
speak of local operations and resources, we mean quantum
operations and resources within each node. On the other hand, we
allow only classical messages to be sent between the nodes. Qubits
belonging to different nodes cannot interact quantum mechanically,
so that no further entanglement can be created between remote
stations. However, qubits
within a node may interact in any way, including with ancillary
(local) resources, and any measurement may be performed on the
components of the node.

\subsubsection{Purification of weakly entangled states}
\label{concepts_entanglement_purification}

\begin{figure}
    \begin{center}
        \includegraphics[height=2.5cm]{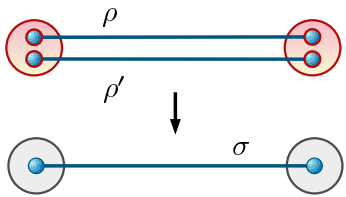}
        \caption{Entanglement purification:
       Each bipartite entangled state $\rho,\rho',\sigma$ is represented by a link.
       The entanglement in two links is
       concentrated by LOCC into a single link, such that,
       by some measure of entanglement, the entanglement
       of the state $\sigma$ is greater than the entanglement of
       $\rho$ and of $\rho'$.}
    \label{fig:purification}
    \end{center}
\end{figure}

We now address the task of concentrating or purifying the
entanglement of two (or more) weakly entangled states into a
pair with higher entanglement, using LOCC only. Suppose that
these states are arranged in a parallel fashion so that
local operations may be performed jointly on all left-hand
members and on all right-hand members as shown in
\fref{fig:purification}. An important task is to find
criteria determining when and how a given set of states can
be transformed into a more highly-entangled target state.

\paragraph{Pure states}

Majorisation theory was developed to answer the question:
What does it mean to say one probability distribution is
more disordered than another?  One application to quantum
mechanics is via the connection between disorder (of the Schmidt coefficients) and
entanglement. Consider an initial pure state $\ket{\alpha}$
and a target state $\ket{\beta}$ in a bipartite
system. These states may have any dimension $d$, that is,
they are pairs of \textit{qudits}. Denoting by $\vec\alpha$
the unit vector of the $d$ Schmidt coefficients of
$\ket{\alpha}$ sorted in decreasing order (and similarly for
$\ket{\beta}$), Nielsen showed in~\cite{Nie99} that a
deterministic LOCC transformation from $\ket{\alpha}$ to
$\ket{\beta}$ is possible if and only if the inequalities
\begin{equation}
    \sum_{i=0}^n \alpha_i \leq \sum_{i=0}^n \beta_i
\end{equation}
hold for all $n\in\{0,\ldots,d-1\}$. In this case, $\vec\alpha$ is
said to be majorised by $\vec\beta$, which is denoted by
$\vec \alpha \prec\vec \beta$; see~\cite{NV01} for more details on
this topic.  Note, for instance, that the maximally entangled state
can be deterministically transformed into any other pure state.
As another example, consider setting
$\rho_0\geq\rho_0'$ for the two pure states $\ket{\rho}$ and $\ket{\rho'}$ depicted
in \fref{fig:purification}, so that one finds $\vec \alpha = (\rho_0\rho_0',\rho_0\rho_1',
\rho_1\rho_0',\rho_1\rho_1')$. In this case, the two pairs of qubits can
be transformed into a single connection $\ket{\sigma}$ if and only if $\rho_0\rho_0'
\leq \sigma_0$.

Moving now to non-deterministic transformations, the optimal probability
for a successful LOCC conversion is~\cite{Vid99}:
\begin{equation}
    \prob(\ket{\alpha}\mapsto\ket{\beta}) = \min_{n}
    \left\{\frac{\sum_{i=n}^{d-1} \alpha_i}{\sum_{i=n}^{d-1} \beta_i}\right\}.
\end{equation}
Using this formula, it is trivial to check that a two-qubit pure state $\ket{\varphi}$
can be transformed into a Bell pair with optimal probability $2\varphi_1$, as
stated in \sref{concepts_entanglement_measures}. Explicitly, this result is obtained by
performing on one of the qubits a generalised measurement defined by the operators
\begin{equation}
    \eqalign{
    M_1 &= \sqrt{\frac{\alpha_1}{\alpha_0}} \ketbra{0}{0} + \ketbra{1}{1},\\
    M_2 &= \sqrt{1-\frac{\alpha_1}{\alpha_0}} \ketbra{0}{0},
    }
    \label{eq:procrustean}
\end{equation}
which is known as the ``Procrustean method'' of entanglement concentration~\cite{BBPS96}.

\paragraph{Mixed states}
The purification of mixed states is a somewhat more difficult problem than that
of pure states. Many techniques have been developed for doing mixed-state
entanglement purification in connection to quantum error-correction~\cite{BDSW96}.
For our purpose, it is sufficient to notice that:
\begin{enumerate}
    \item In contrast with the case of pure states, at least two
          copies of a Werner state are needed to get, by LOCC and with
          finite probability, a state of higher fidelity~\cite{LMP98}.

    \item Perfect Bell pairs can be obtained from $N$ Werner states only in the
        limit $N\rightarrow\infty$~\cite{Ken98}.
\end{enumerate}
The first purification scheme, which was proposed by Bennett \etal in~\cite{BBP+96},
is depicted in \fref{fig:purification}: the two states $\rho=\rhoW(x)$ and
$\rho'=\rhoW(x')$ are purified into the state $\sigma = \rhoW(x'')$, with
\begin{equation}
    x'' = \frac{x+x'+4 x x'}{3 + 3 x x'}.
    \label{eq:purifyW}
\end{equation}
The resulting state is closer to the target state $\ket{\Phi^+}$ if
both $\rho$ and $\rho'$ are entangled (that is, if $x,x'>1/3$) and if
$x>x'>2x/(1+4x-3x^2)$. It is important to note that this operation is
not deterministic since it succeeds only with probability
\begin{equation}
    \prob(\rho\otimes\rho'\mapsto\sigma) = \frac{1+x x'}{2}.
    \label{eq:purifyW_prob}
\end{equation}
This procedure can be iterated, with $x$ increasing after each step,
until it is arbitrarily close to $1$ (considering perfect operations);
this asymptotic technique is sometime referred as {\em distillation}.
However, one is often interested in the yield, which is defined as the
asymptotic ratio of the number of input states to the number of output
states. The above protocol requires a diverging number of states to
produce one arbitrarily pure state, and thus has a vanishing yield. In
order to get a positive yield, the previous method is applied until
states of sufficiently large $x$ are generated, and then one switches
to purification techniques using one-way communication. Many
improvements and variants over this construction exist;
see~\cite{HHHH09,DB07} and references therein.

\subsubsection{Entanglement swapping}
\label{concepts_QN_swap}

\begin{figure}
    \begin{center}
        \includegraphics[height=2.7cm]{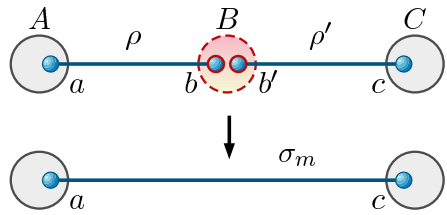}
        \caption{Entanglement swapping: the middle station performs a measurement
        in the Bell basis on its qubits. This creates a quantum connection
        between the two opposite stations, which depends on the outcome $m$ of
        the measurement.}
    \label{fig:swapping}
    \end{center}
\end{figure}

The basic operation to propagate the entanglement in a quantum network is
the so-called \textit{entanglement swapping}~\cite{BBC+93,BPM+97,ZZHE93,BVK98}
depicted in \fref{fig:swapping}: By performing
a Bell measurement on the qubits $b$ and $b'$ at the station $B$, one creates
a quantum link between the previously unconnected stations $A$ and $C$. Then, a local
unitary that depends on the outcome $m$ of the measurement is applied
on the qubit $c$, so that the resulting entangled pair has the standard form
given in \eref{eq:phi} or in \eref{eq:Wx}. This operation is equivalent to
teleporting $b$ to $c$.

For two mixed states $\rho=\rhoW(x)$ and $\rho'=\rhoW(x')$, it is easy to see
that the entanglement swapping produces the quantum state $\sigma_m = \rhoW(x x')$
for all measurement outcomes, that is,
\begin{equation}
    C(\sigma_m) = \max\left\{0,\frac{3xx'-1}{2}\right\} \quad\forall \  m.
        \label{eq:Werner_swap_C}
\end{equation}

In the case of pure states $\ket{\rho}$ and $\ket{\rho'}$, however, the result
depends on the outcome, and one gets either a Bell pair
or a state that is weakly entangled. Remarkably, the average entanglement of
the resulting states $\ket{\sigma_m}$ is not less than that of the
initial states:
\begin{equation}
    \bar{E}(\sigma_m) = \frac{1}{4} \sum_{m=1}^{4} E(\sigma_m) =
    2 \min\{\rho_1,\rho_1'\},
\end{equation}
which is, for $\ket{\rho}=\ket{\rho'}$, the ``conserved entanglement'' described in~\cite{BVK99}.
This property of pure states will be used in the various
protocols of entanglement propagation described in \sref{percolation}.

Maximising the average entanglement of the outcomes is of prime importance
for random or statistical processes. 
However, one may also desire
that every possible outcome results in a state with a reasonable amount of entanglement.
In this scenario one can use the rotated Bell basis $\cB_X
\equiv (X \otimes \id_2)\,\cB$ to perform the entanglement swapping;
see section 1.1 in~\cite{Per10b}. In this case, one gets four outcomes
$\ket{\sigma_m}$ satisfying
\numparts
\begin{equation}
    E(\sigma_m) = 1 - \sqrt{1-16 \rho_0\rho_1\rho_0'\rho_1'}
    \quad\forall m,
    \label{eq:intro_swap_pureXE}
\end{equation}
or, equivalently,
\begin{equation}
    C(\sigma_m) = C(\rho)\,C(\rho') \quad\forall m.
    \label{eq:intro_swap_pureXC}
\end{equation}
\endnumparts

The entanglement swapping procedure can be iterated, creating a quantum
connection between qubits that are more and more distant. However, the resulting
long-distance entanglement that is generated in this way decreases exponentially
with the number of swappings. This is clear in the case of mixed states and for
pure-state Bell measurements in the $\cB_X$ basis; the general proof can
be found in section 1.3 in~\cite{Per10b}. Because of this important loss of entanglement,
new schemes have to be designed to efficiently entangle any two stations of
a quantum network. The various protocols proposed so far that achieve this
task are described in the following chapters.

\paragraph{Noisy operations}

Thus far, we have assumed that all quantum operations are ideal or perfect.
However, in practice, every operation or measurement introduces
some noise. Here we consider two sources of error: those
arising from applying an imperfect gate (unitary operation), and those
arising from an imperfect measurement. In the first case we
model the error by including a small depolarising term to a channel, which
replaces a fraction of the density operator with a completely decoherent
(\ie\ unknown) state~\cite{BDCZ98,Per10b}.
For a multi-qubit state, a gate $O_{S}^{\rm{ideal}}$ acting on a subset $S$ of qubits
with an error probability $\varepsilon$ is replaced by the map
\begin{equation}
 \label{eq:noisy_gate}
 \rho \mapsto O_{S}[\rho] = (1-\varepsilon)\ O_{S}^{\rm{ideal}}[\rho] + \varepsilon'\ 
    \id_{S} \otimes \tr_{S}[{\rho}],
\end{equation}
where $\tr_{S}$ denotes the partial trace over the subsystem $S$, and
$\varepsilon'$ is such that the resulting state has trace one.
It can be shown that this model correctly models isotropic errors for
single qubit rotations~\cite{DBCZ99}.
On the other hand, suppose we model a measurement error on a single qubit
by assuming that we have a small fixed probability $\varepsilon$ of reading
$1$ when the qubit was actually measured into the $0$ state. In
this case, the measurement operators read:
\begin{equation}
    \eqalign{
    M_0^\eta & = \sqrt{1-\varepsilon}\ketbra{0}{0} + \sqrt{\varepsilon}\ketbra{1}{1}, \\
	M_1^\eta & = \sqrt{1-\varepsilon}\ketbra{1}{1} + \sqrt{\varepsilon}\ketbra{0}{0}.
    }
\end{equation}
Propagating the errors according to this simple model
allows us to estimate the error resulting from a particular
protocol.
\subsection{The quantum repeaters}
\label{concepts_repeaters}

\begin{figure}
    \centering
    \subfloat[]{\label{fig:repeater_a}
        \includegraphics[width=.45\linewidth]{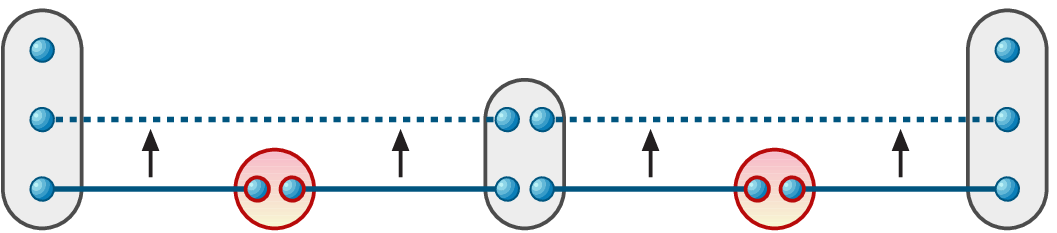}}
    \qquad\quad
    \subfloat[]{\label{fig:repeater_b}
        \includegraphics[width=.45\linewidth]{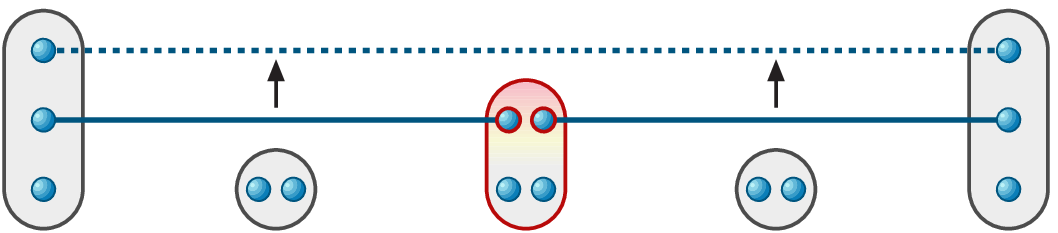}}
    \caption[Quantum repeaters]{%
    The nested purification protocol used by the quantum repeaters.
    (a) Elementary connections are continuously
    generated between neighbouring stations, and an entanglement swapping is
    performed at every second node. The resulting states have less entanglement,
    but they are repeatedly purified.
    (b) Once the states have regained a sufficiently large fidelity, the procedure is
    iterated at a higher level. This eventually leads to an entangled pair
    of qubits between the two endpoints of the chain.}
    \label{fig:repeater}
\end{figure}

A great deal of theoretical and experimental effort
has been put forth to distribute entangled states over
long distances using essentially one-dimensional
lattices. As we mentioned above, networks were introduced
to solve problems caused by unavoidable loss and
decoherence through free-space and fibre links. In
particular, quantum repeaters, which have entanglement
swapping at their heart, have received the most attention.

The initial proposals for quantum repeaters use a
hierarchical scheme of swapping and purification 
steps~\cite{CZKM97,BDCZ98,DBCZ99}.  Including purification in the
protocol is necessary once one introduces real-world noise
and errors.  Noise enters the system in two ways. First,
each operation or measurement reduces the fidelity of
the desired state. This noise is modelled
as described in the previous section.
Second, even in the case of perfect operations, if one
begins with a slightly impure state, then the state
resulting from a protocol decays rapidly in the number of
operations, such as the swapping in \eref{eq:Werner_swap_C},
to a useless separable state. As explained above, the losses
suffered by a state transmitted through the links of the
network increases exponentially with distance. This
introduces a maximum length of elementary links because
there is a minimum fidelity $F_{\min}$ \eref{eq:Fmin} below
which purification is impossible. The repeater protocol
first prepares several states of $F>F_{\min}$ along a
relatively short link and stores them in quantum
memories. Then they are used to produce a pair with
higher fidelity through entanglement 
purification. Entanglement swapping is then performed on
two of these purified states on neighbouring links thus
creating entanglement across a distance that is twice the
length of the elementary link. The noisy swapping again
reduces the fidelity, so that more quantum memories and more
purifications is needed. This procedure is iterated so
that, in principle, highly entangled states can be created
across long distances.

Based on the initial protocol for quantum repeaters, many
improvements have been suggested; see~\cite{SSRG11} for a review
on this topic. For instance, it has been shown
that the number of qubits per station does not have to grow with
the distance~\cite{CTSL05}. Yet, the realisation of quantum repeaters
is still a very challenging task, which is mainly due to the need of
reliable quantum memories~\cite{HKBD07}.

\subsubsection{Implementations}
\label{concepts_entanglement_implementation}

All building blocks needed to construct a quantum network have been
demonstrated, and, in fact, small-scale quantum networks are now a
reality. For instance, while the first demonstrations of teleportation
were made in laboratory scales ~\cite{BPM+97,BBMHP98,FSBFKP98},
presently, entangled photons distributed in free space can be used for
teleportation over $150$~km~\cite{MNH+12,YRL+12} (see also the recent
improvements in the direction of telecommunication~\cite{MRTZG03}).
Moreover atom-photon interfaces have also been used in the
demonstration of teleportation~\cite{BCS+04,SKOJHCP06} and
entanglement swapping was also shown in different
scenarios \cite{PBWZ98,HBGSSZ07}.

One of the first implementations of a quantum network, the DARPA
Quantum Network, consists of several nodes and supports both
fibre-optic and free-space links~\cite{Ell04}. It is capable of
distributing quantum keys between sites separated by a few tens of
kilometres.  Another example is the SECOCQ network in Vienna, which
consists of six nodes connected by links ranging from 6 to 85
km~\cite{SECOQC09}.


%% file: sections/percolation.tex
\newpage
\setcounter{footnote}{0}
\section{Entanglement percolation}
\markboth{\sc Entanglement percolation}{}
\label{percolation}

An approach to entangling distant parties that is conceptually
 different from the quantum repeater was proposed in~\cite{ACL07}.  In
 this paper, the main question is: given a quantum network,
or graph, which operations should be performed on the nodes so that
 entanglement is best propagated?\\
A first answer is presented in \sref{percolation_deterministic}:
for some graphs, the gain obtained from purification of weakly
entangled states can balance, or even surpass, the loss of
entanglement resulting from the swappings. This result is
promising, but it is merely an adaptation of the quantum repeater
protocol to specific graphs and quantum states. In fact, it
somewhat replaces the repeated generation of elementary links by a
deterministic accumulation of the entanglement of existing links,
provided that they are in a pure state, 
or that the graph has a specific structure.\\
The solution proposed in~\cite{ACL07} is of a different nature.
Its underlying idea is that in one-dimensional networks, any
defective link destroys the whole procedure, while in higher-dimensional
networks the information can still reach its destination through other paths.
This phenomenon is related to percolation theory, which states that if there
are not too many defective links in, say, a square lattice, then any two nodes
of the network are connected by a path with non-vanishing probability.
Strategies based on this idea are described in \sref{percolation_pure}. Initially
limited to pure states, they have since been generalised to some special cases
of mixed states; see \sref{percolation_mixed}.

\subsection{Deterministic protocols based on purification}
\label{percolation_deterministic}

In this section, we show that qubits can become entangled over large scales
in some two-dimensional (planar) graphs, using predetermined sequences of entanglement
swappings and purifications. Very close to the quantum repeaters in spirit,
this method can be applied either with mixed states in a restricted class of graphs
(\sref{percolation_det_hierarchy}) or with pure states in lattices of high
connectivity (\sref{percolation_det_regular}).

\subsubsection{Hierarchical graphs}
\label{percolation_det_hierarchy}

\begin{figure}
    \centering
    \subfloat{\label{fig:hierarchy_a}
        \includegraphics[width=4.7cm]{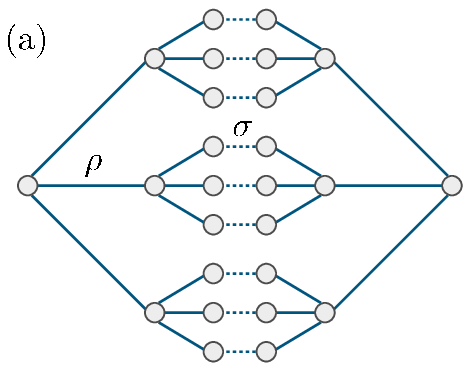}}
    \hspace{.05\linewidth}
    \subfloat{\label{fig:hierarchy_b}
        \includegraphics[width=4.7cm]{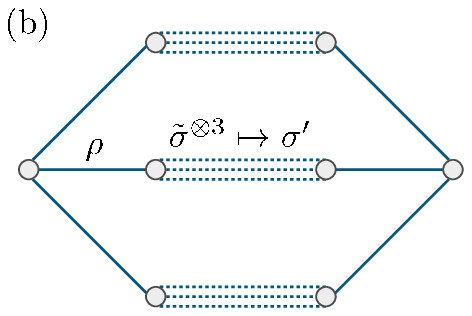}}
    \caption[Two joined ternary trees]{%
        A hierarchical graph that consists of two infinite ternary trees.
        The links of the trees are in an entangled state $\rho$, while
        the central connections may be in a different state $\sigma$.
        (a) Entanglement swapping is performed at each extremity of the
        trees, creating three links $\tilde{\sigma}$ between every pair of
        new leaves.
        (b) Every group of three states $\tilde{\sigma}$ is purified into
        one connection $\sigma'$. The process reaches any level of the hierarchical
        graph if the entanglement of $\sigma'$ is greater than
        or equal to that of $\sigma$.}
    \label{fig:hierarchy}
\end{figure}

Hierarchical graphs iterate certain geometric
structures, so that at each level of iteration either the number of neighbours
or the length of the connections increases. Various such graphs were
considered in~\cite{PCA+08}; let us give here yet another example in which
both pure and mixed state entanglement can be generated between nodes that lie at
any level of the hierarchy. In this example, two infinite ternary trees, in
which each link is an entangled state $\rho$, are connected at their leaves by
a state $\sigma$; see \fref{fig:hierarchy_a}. The protocol runs as follows:
First, two entanglement swappings are performed on the central links $\rho$,
$\sigma$, and $\rho$, yielding a state $\tilde{\sigma}$. Second, the three states
$\tilde{\sigma}$ that connect the new leaves of the ternary trees are purified,
leading to a single connection $\sigma'$. The procedure can be iterated as
long as $E(\sigma')\geq E(\sigma)$. In this
way, nodes lying at higher and higher level of the hierarchical graphs
become entangled, that is, larger and larger quantum connections are created.
In the following, we determine the minimum amount of pure or mixed-state
entanglement of $\rho$ and $\sigma$ for this strategy to be successful.

\begin{figure}
    \begin{center}
        \includegraphics[width=4.7cm]{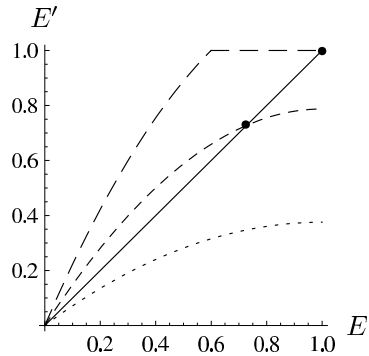}
        \caption{Graph of the recursion relation \eref{eq:purification_recPure}.
        Three regimes have to be distinguished: $\mu<\mu_c$ (dotted bottom curve),
        $\mu_c<\mu<\mu^*$ (dashed middle curve), and $\mu>\mu^*$ (long-dashed
        upper curve). A non-trivial fixed point (bullet) appears only if $\mu>\mu_c$,
        and a Bell pair is created after a finite number of iterations if
        $\mu>\mu^*$.}
    \label{fig:hierarchy-rec}
    \end{center}
\end{figure}

\paragraph{Pure states}
In \sref{concepts_QN_swap}, we saw that the result of the
entanglement swappings depends on the basis that is chosen
for performing the two-qubit measurement. In order to
facilitate the comprehension of the mechanism, we choose the
$\cB_X$ basis, so that all outcomes are equally entangled:
$C(\tilde{\sigma}) = C^2(\rho)\,C(\sigma)$; see
\eref{eq:intro_swap_pureXC}.  The optimal purification of
$\tilde{\sigma}^{\otimes 3}$ into $\sigma'$, as described in
\sref{concepts_entanglement_purification}, requires working
with the Schmidt coefficient $\tilde{\sigma}_0 = (1 +
\sqrt{1-C^2(\tilde{\sigma})})/2$.  One then finds $\sigma_0'
= \max\{\frac{1}{2},\tilde{\sigma}_0^3\}$, and the recursion
relation for the entanglement of the central connections
reads:
\begin{equation}
    E' =  \min\left\{1,\,2-\frac{1}{4}\Big(
    1+\sqrt{1-\mu\, E(2-E)}\Big)^3\right\},
    \label{eq:purification_recPure}
\end{equation}
with $E'=E(\sigma')$ and $E=E(\sigma)$. In this equation, the parameter $\mu = C^4(\rho)$
lies in the interval $[0,1]$. It is an easy calculation to show that $E'<E$ for any value of
$E$ if $\mu<\mu_c=\frac{1}{3}$, that is, if $E(\rho)<E_c\approx 0.35$.
This means that the entanglement cannot be propagated in the hierarchical graph
if the links $\rho$ are too weakly entangled. In contrast, if $\mu>\mu_c$, one
stable and non-trivial fixed point appears in \eref{eq:purification_recPure};
see \fref{fig:hierarchy-rec}. In this case, nodes lying at any level of the
hierarchy can be connected by an entangled state of two qubits;
one further shows that this state is a perfect Bell pair if
$\mu\geq\mu^*\approx 0.655$.

\paragraph{Mixed states}
The scenario of a mixed-state hierarchical graph, where the connections are
Werner states $\rho = \rhoW(x)$ and $\sigma = \rhoW(y)$, is quite similar to
that of pure states. First, two consecutive entanglement swappings are
performed on the central states $\rho$, $\sigma$, and $\rho$, leading to a
state $\tilde{\sigma} = \rhoW(\tilde{y})$ with $\tilde{y} = x^2y$. Then, we
try to concentrate the entanglement of the three states $\tilde{\sigma}$ into
one connection $\sigma'$. As for the pure states, we would like this operation
to be deterministic, but the purification of mixed states is intrinsically
probabilistic. In order to get a result in a predictable fashion, we
purify two connections only and take the third one if the purification
failed. From \eref{eq:purifyW} and \eref{eq:purifyW_prob}, 
this succeeds with probability $p = (1+\tilde{y}^2)/2$ and the average
entanglement of $\sigma'\equiv\rhoW(y')$ is
\begin{equation}
    y'(x,y) = \frac{x^2y}{6}(5+4x^2y-3x^4y^2).
      \label{eq:purification_recMixed}
\end{equation}
If the links of the graph satisfy $x>x_c=\sqrt{18/19}$ and $y>y_c(x)=
(2x-\sqrt{19x^2-18})/(3x^3)$, then a stable fixed point $y^*$ appears. In this
case, iterating the entanglement swappings and the purifications generates
some long-distance pairs of qubits whose entanglement approaches $y^* =
(2x+\sqrt{19x^2-18})/(3x^3)$.

\subsubsection{Regular graphs}
\label{percolation_det_regular}

\begin{figure}
    \begin{center}
        \includegraphics[width=4.7cm]{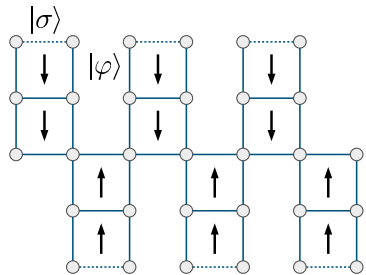}
        \caption{A ``centipede'' in the square lattice: the entanglement of the
        links is progressively concentrated along the ``spine'' of the centipede.
        If the amount of entanglement present in the links is larger than a
        critical value $E_c$, then perfect Bell pairs are obtained on the
        central path and long-distance entanglement can be generated.}
    \label{fig:centipede}
    \end{center}
\end{figure}

We have just shown that entangled pairs of qubits can be generated over a large
distance in graphs with a hierarchical structure, under the condition that the
entanglement of the bonds if larger than a critical value $E_c$. The self-similarity
of these graphs allows one to design natural sequences of entanglement swappings
and purifications but suffers a physical limitation: either the length of the
bonds or the number of qubits per node grows exponentially with the iteration depth.
We now consider regular two-dimensional lattices, that is, periodic
configurations of links throughout the plane in which the nodes have a fixed
number $Z$ of neighbours.\\

A deterministic strategy to entangle two infinitely distant nodes
in lattices in which each node has a number of nearest neighbours
 $Z\geq4$ was proposed in~\cite{PCA+08}.
It is very similar to the recursive method developed in the previous example,
but it works only with pure states. In the case of a square lattice of
links $\ket{\varphi}$, the idea is to sequentially shorten the legs of a
``centipede'', so that the entanglement of the links is gradually concentrated;
see \fref{fig:centipede}. This eventually yields a perfect Bell pair at the spine
of the centipede, on which infinitely many entanglement swappings can be applied,
and therefore long-distance entangled pairs of qubits are generated. More
precisely, one starts by applying two entanglement swappings in the $\cB_X$
basis on the states $\ket{\sigma}$ (dotted line in \fref{fig:centipede}) 
and its neighbour states $\ket{\varphi}$ at the
extremity of each leg. This results in a state $\ket{\tilde{\sigma}}$, and as in
the previous example, we have $C(\tilde{\sigma}) = C^2(\varphi)\ C(\sigma)$.
The difference is that the purification is now performed on $\ket{\tilde{\sigma}} \otimes
\ket{\varphi}$ rather than on $\ket{\tilde{\sigma}}^{\otimes 3}$. Very similarly
to \eref{eq:purification_recPure}, one finds the following recursion relation:
\begin{equation}
    E' =  \min\left\{1,\,2-\varphi_0\Big(
    1-\sqrt{1-\mu\, E(2-E)}\Big)\right\},
    \label{eq:purification_centipede}
\end{equation}
where $\mu = C^4(\varphi)$ is a function of $\varphi_0$. One can show
that there always exists a non-trivial stable fixed point for this equation.
However, the fixed point of \eref{eq:purification_centipede} is strictly smaller
than unity when $E(\varphi)<E_c \approx 0.649$. In this case, although we do
concentrate some entanglement along the spine of the centipede, we still face
the problem that the spine is a one-dimensional system, which therefore exhibits
an exponential decrease of the entanglement with its length. On the other hand,
if $E(\varphi)>E_c$, then the fixed point is reached in a finite number of steps,
and a maximally entangled state is generated. Since the spine now consists of perfect
connections, any two nodes lying on it can share a Bell pair, regardless of their
distance.

\subsection{Percolation of partially entangled pure states}
\label{percolation_pure}

We have demonstrated that a way to generate long-distance
entanglement in a lattice is to purify a ``backbone'' of
Bell pairs and then to perform some entanglement swappings
along this path.  Three conditions have to be satisfied for
this method to work.  First, the nodes that belong to the
backbone must have at least four neighbours each: two
connections are part of the backbone, while the other two
are used to purify the former.  Second, the entanglement of
the bonds has to be larger than a critical value $E_c$ that
depends on the lattice geometry.  Third, the links have to
be pure states and not mixed states, because the
purification of a finite number of Werner states never leads
to a Bell pair (\sref{concepts_entanglement_purification}).
Since this deterministic strategy creates a chain of Bell
pairs by using only a strip of finite width from the
lattice, it seems that it does not exploit the full
potential of two-dimensional networks.  In this section,
we review the method of entanglement percolation%
\footnote{%
  Note that the probabilistic nature of quantum physics
  makes percolation theory a particularly well-adapted
  toolbox for the study of quantum systems which undergo,
  for instance, measurements. Ideas of percolation theory
  are for example useful in the context of quantum computing
  with non-deterministic quantum gates~\cite{KRE07}. We
  refer the interested reader to pp.~287--319 in~\cite{SBC09}
  for an overview of the application of
  percolation methods to the field of quantum information.}
that was published in~\cite{ACL07} and that partially relaxes the above
conditions:
\begin{enumerate}
    \item percolation is a genuine two- (or multi-) dimensional phenomenon,
    and thus it applies to any lattice;
    \item entanglement percolation undergoes a phase transition with
    the entanglement of the links, but the corresponding critical value is
    smaller than that of the purification method~\cite{ACL07,PCA+08,LWL09};
    \item in certain lattices, this method applies also to two-qubit mixed states
    of rank less than four~\cite{BDJ09,BDJ10}.
\end{enumerate}

The simplest application of entanglement percolation in
infinite lattices is presented in
\sref{percolation_pure_CEP}. In this case, the connection to
classical percolation theory is straightforward and
entanglement thresholds are readily determined.  Then, we
show that modified versions of this \textit{classical
  entanglement percolation} (CEP) yield lower
thresholds. For instance, we reconsider the hexagonal lattice with
double bonds proposed in~\cite{ACL07}, in which quantum
measurements lead to a local reduction of the entanglement but
change the geometry of the lattice. This operation increases the
connectivity of the graph and lowers the entanglement threshold,
an effect which is called \textit{quantum entanglement
percolation} (QEP); see \sref{percolation_pure_QEP}. Finally, an alternative
construction that uses multipartite entanglement is given in
\sref{percolation_pure_MEP}. Using incomplete measurements, this
protocol creates entangled states of more than two qubits and
improves not only the entanglement threshold, but also the success
probability of the protocol for any amount of entanglement in the
connections for all the cases considered in
\sref{percolation_pure_MEP}.

\subsubsection{Classical entanglement percolation}
\label{percolation_pure_CEP}

\begin{figure}
    \centering
    \subfloat[][$p=0.25$]{\label{fig:bondpercolation_a}
        \includegraphics[width=.28\linewidth]{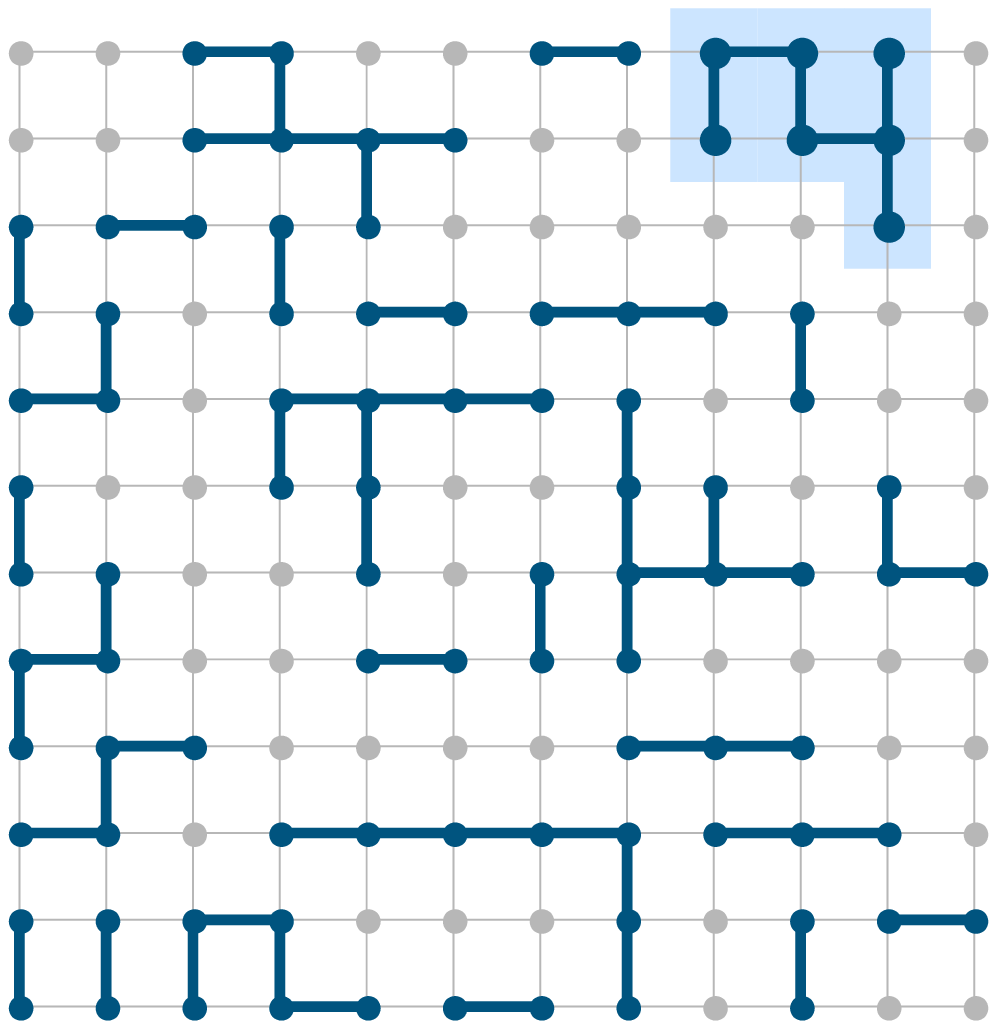}}
    \qquad
    \subfloat[][$p=p_c=0.5$]{\label{fig:bondpercolation_b}
        \includegraphics[width=.28\linewidth]{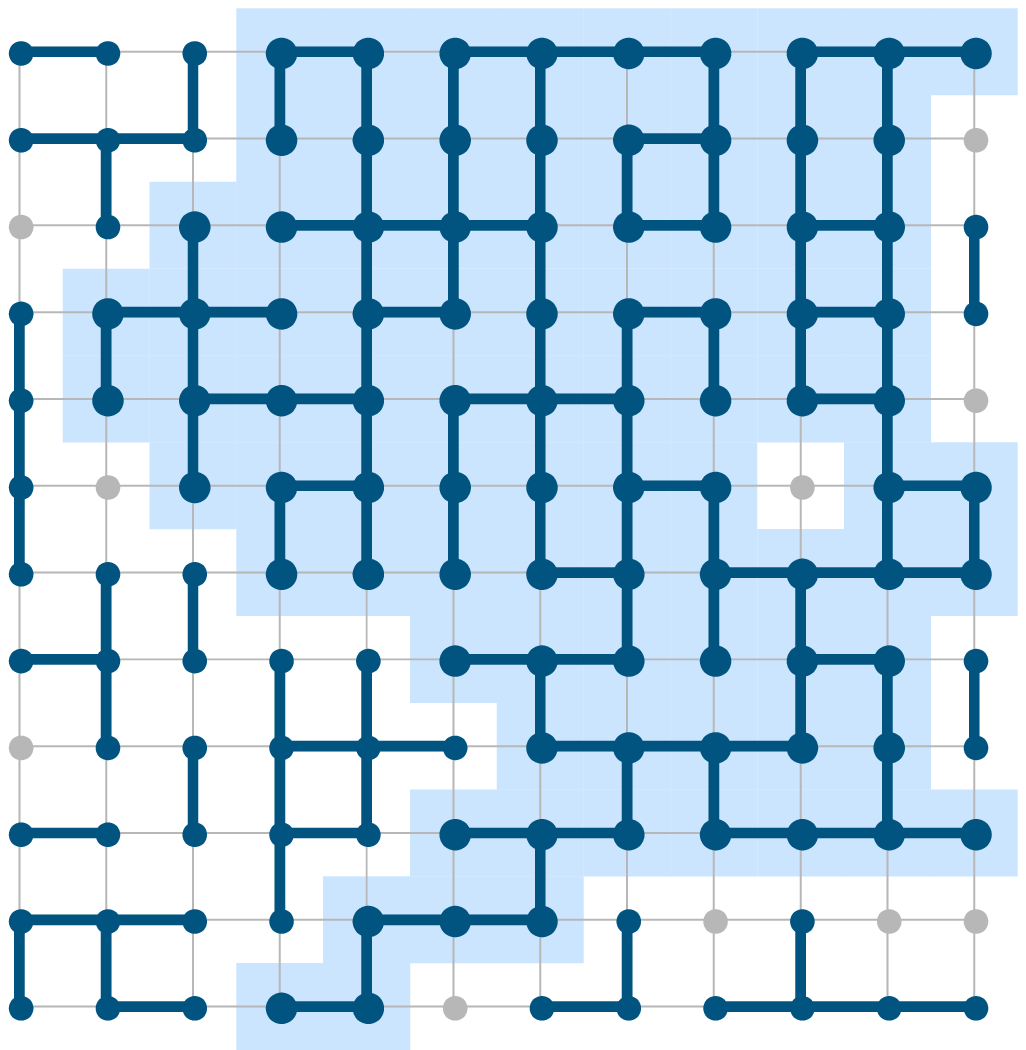}}
    \qquad
    \subfloat[][$p=0.75$]{\label{fig:bondpercolation_c}
        \includegraphics[width=.28\linewidth]{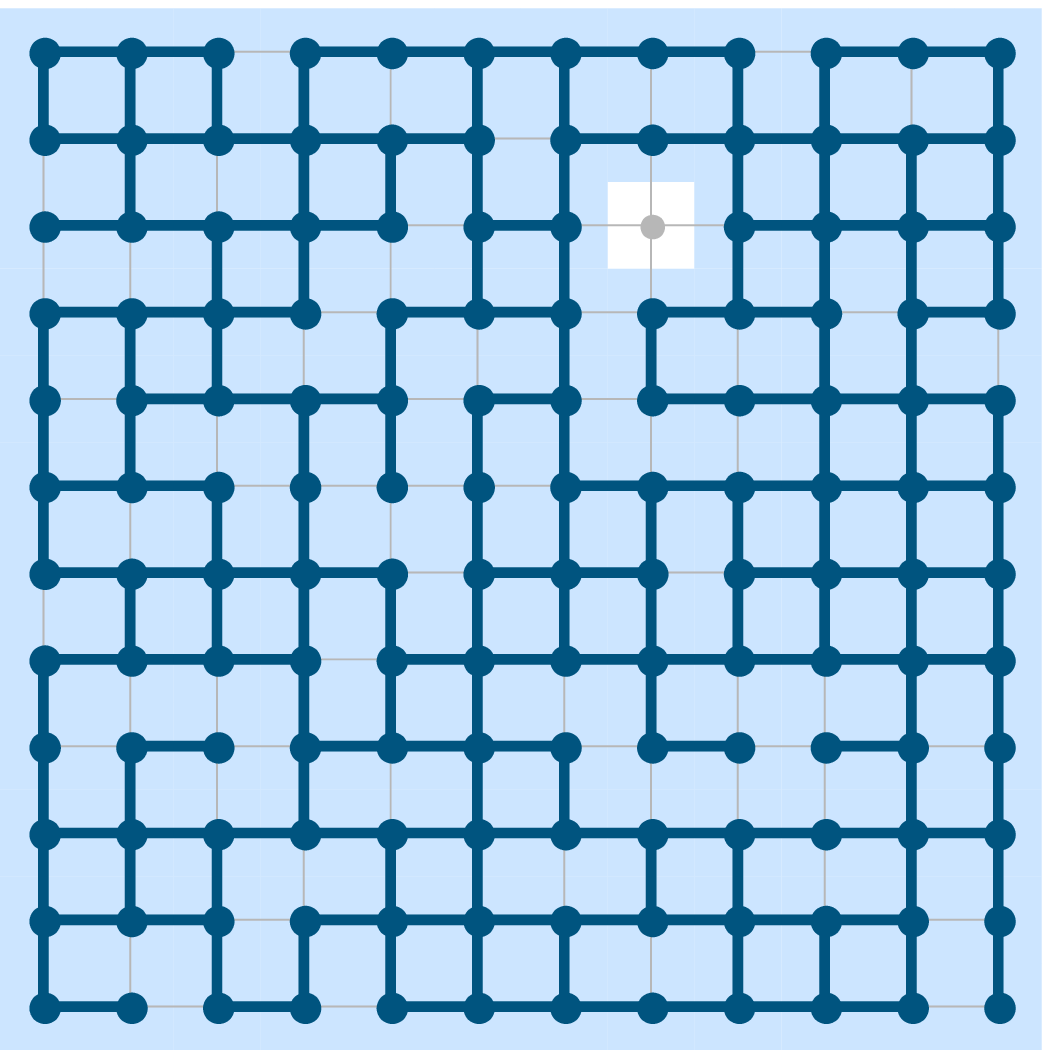}}
    \caption[Bond percolation]{%
    In classical bond percolation, the bonds are open (present) with probability
    $p$ and closed otherwise. Groups of nodes connected by open bonds are
    called clusters. In these examples, the largest cluster is highlighted:
    typically, it is small and bounded for $p<p_c$, but it spans a finite
    proportion of the lattice for $p>p_c$. In the latter case,
    there is a unique infinite cluster $\cluster$.}
    \label{fig:bondpercolation}
\end{figure}

Classical percolation is perhaps the fundamental example of
critical phenomena, since it is a purely statistical
one~\cite{Gri99}. At the same time, it is quite universal
because it describes a variety of processes, with
applications in physics, biology, ecology, engineering,
etc.~\cite{SA91}.  Two types of models are typical
considered, site- and bond percolation.  In bond
percolation, the neighbouring nodes of a lattice are
connected by an \textit{open bond} with probability $p$,
whereas they are left unconnected with probability $1-p$;
see \fref{fig:bondpercolation}. In site percolation, the
sites ({\it i.e.} nodes) rather than bonds are occupied with
probability $p$. In either case, for an infinite lattice,
one would like to know whether an infinite open cluster
exists, that is, whether there is a infinitely long path of
connected nodes.  It turns out that an unique infinite
cluster $\cluster$ appears if, and only if, the connection
probability $p$ is larger than a critical value $p_c$ that
depends on the lattice. Few lattices have a threshold that
is exactly known. Among them, we find the important
honeycomb, square, and triangular lattices, with
$p_c^{\hexagon}= 1-2\sin(\pi/18)$, $p_c^{\Box} = 1/2$,
and $p_c^{\vartriangle} = 2\sin(\pi/18)$, respectively~\cite{SE64}.\\

Suppose now that we want to generate some entanglement between two distant stations
$A$ and $B$ in a quantum lattice, where each connection denotes
a partially-entangled pure state $\ket{\varphi}$. Classical entanglement
percolation (CEP) runs as follows~\cite{ACL07}: every pair of neighbouring nodes tries
to convert its two-qubit state into a Bell pair, which succeeds
with an optimal probability $p=E(\varphi)$; see \sref{concepts_entanglement_purification}.
If this value is larger than the threshold $p_c$ of the lattice, that is, if
the entanglement of the links is large enough,
then an infinite cluster $\cluster$ appears. The probability that both $A$ and
$B$ belong to $\cluster$ is strictly positive, and in this case a path of Bell pairs
between these two nodes can be found. Then, exactly as described in the previous
section, one performs the required entanglement swappings along this path such
that $A$ and $B$ become entangled. Note that the path of Bell pairs is randomly
generated by the measurement outcomes at the nodes, which contrasts with the
deterministic location of the ``backbone'' generated by the purification method.\\

A quantity of primary interest when studying the efficiency of CEP is the
\textit{percolation probability}
\begin{equation}
    \theta(p)\equiv P(A\in\cluster),
\end{equation}
which is the probability that a node $A$ belongs to the infinite cluster. This value
is closely related to the percolation threshold: in fact, in an infinite lattice
we have $\theta(p)=0$ for $p<p_c$, whereas $\theta(p)>0$ for $p>p_c$.
In our case, we are interested in the probability $P(A\connected B)$ of creating
a Bell pair between two nodes $A$ and $B$ separated by a distance $L$. For $p<p_c$,
this probability decays exponentially with $L/\xi(p)$~\cite{BR06}, where the
\textit{correlation length} $\xi(p)$ describes the typical radius of an open
cluster. Above the critical point, the two nodes are connected only if they
are both in $\cluster$. In the limit of large $L$, the events $\{A\in\cluster\}$
and $\{B\in\cluster\}$ are independent, so that
\begin{equation}
    P(A\connected B) = \theta^2(p),
\end{equation}
and therefore the problem is reduced to studying $\theta(p)$.

\subsubsection{Quantum entanglement percolation}
\label{percolation_pure_QEP}

A natural question is whether the entanglement thresholds $E_c=p_c$
defined by the classical percolation theory are optimal. In fact,
percolation of entanglement represents a related but different theoretical problem,
where new bounds may be obtained. This is of course equivalent to
determining if the measurement strategy based on local Bell pair conversions
is optimal in the asymptotic regime. Several examples that go beyond the classical
picture were developed in~\cite{ACL07,PCA+08,LWL09}, proving that CEP is not optimal;
such results are referred to as ``quantum entanglement percolation'' strategies.
All these examples are based on the average conservation of the entanglement after
one swapping, as described
in \sref{concepts_QN_swap}, but they do not provide a general construction
to surpass the classical percolation strategy. In the following paragraph we
review the original example of~\cite{ACL07}, since it gives some insights into
the way a quantum lattice can be transformed by local measurements, and we let
the reader consult the articles~\cite{PCA+08,LWL09} for the other examples.
Finally, an improved strategy that makes use of
multipartite entanglement will be described in \sref{percolation_pure_MEP}.

\begin{figure}
    \centering
    \subfloat[][]{\label{fig:percolation_honeycomb_a}
        \includegraphics[height=3.8cm]{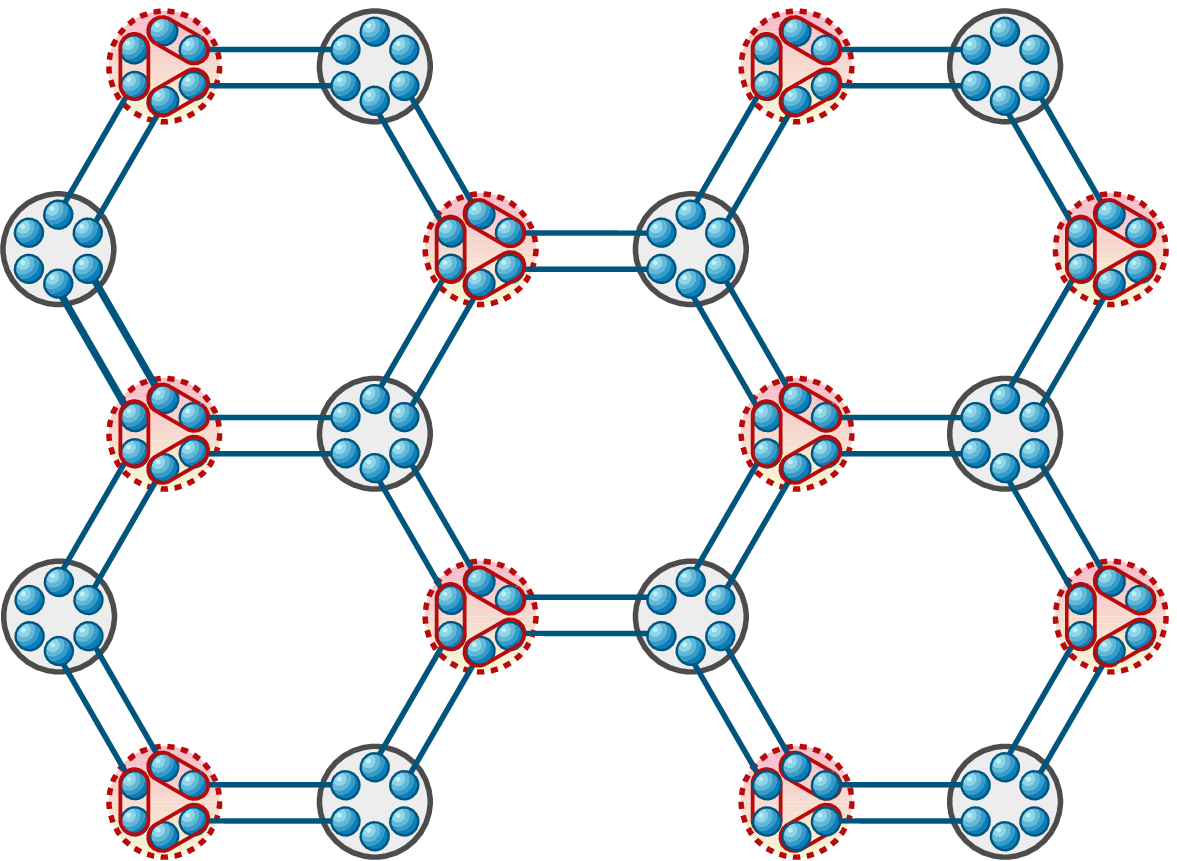}}
    \qquad\quad
    \subfloat[][]{\label{fig:percolation_honeycomb_b}
        \includegraphics[height=3.8cm]{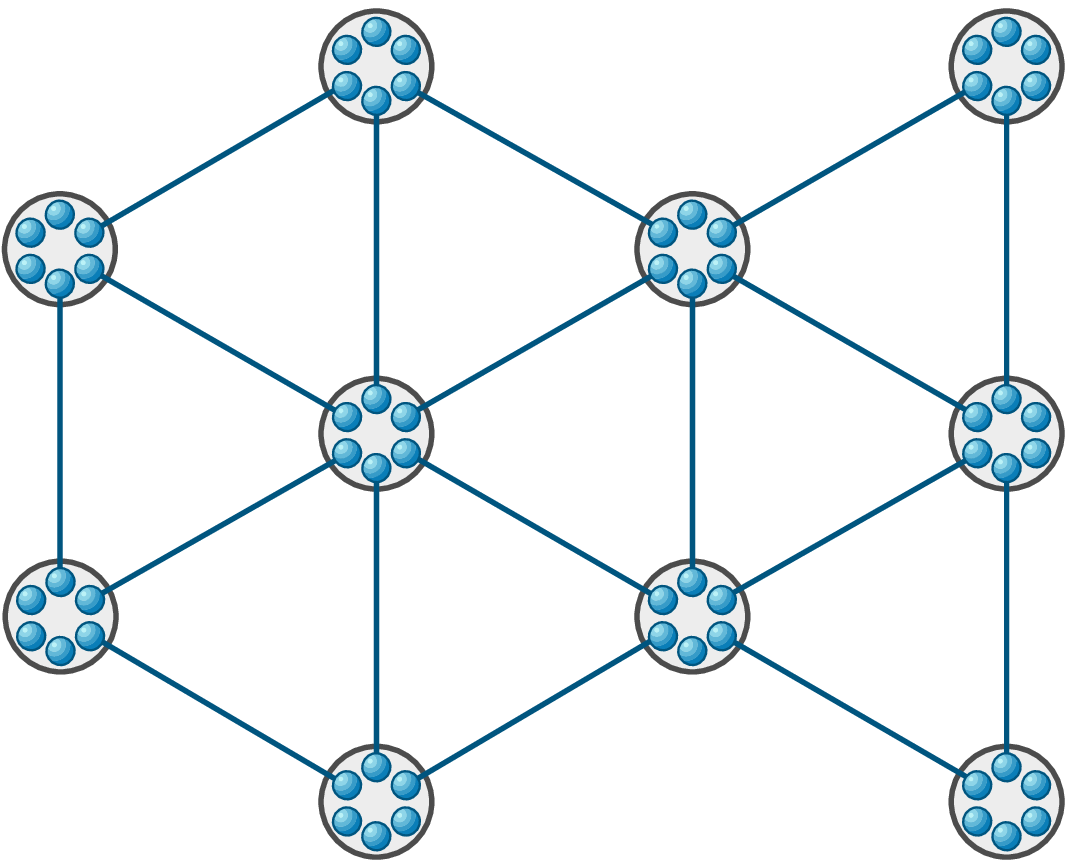}}
    \caption[Honeycomb lattice with double bonds]{%
    Each bond of the honeycomb lattice consists of two copies of the state
    $\ket{\varphi}$.
    (a) The dashed nodes perform three entanglement swappings in the Bell basis
    $\cB$.
    (b) A triangular lattice of identical average entanglement is obtained, and
    CEP is now possible in the new lattice.}
    \label{fig:percolation_honeycomb}
\end{figure}

\paragraph{Honeycomb lattice with double bonds}
Let us consider a honeycomb lattice where each pair of neighbouring nodes is
connected by two copies of the same state $\ket{\varphi}$; see
\fref{fig:percolation_honeycomb_a}. The CEP protocol converts all bonds shared
by two parties into a single connection, and from majorization theory we know
that a double bond can be optimally purified into one pair of qubits with
entanglement $E' = 2(1-\varphi_0^2)$. Setting this value to be equal to the
percolation threshold for the honeycomb lattice, one finds that the entanglement
can be propagated if $E(\varphi)>E_c$, with
\begin{equation}
    E_c = 2\left(1-\sqrt{1-\frac{p_c^{\hexagon}}{2}}\right)\approx 0.358.
\end{equation}
However, there exists another measurement pattern yielding a better percolation
threshold (\fref{fig:percolation_honeycomb}): half the nodes perform on their
qubits three entanglement swappings, which maps the honeycomb lattice into a triangular
one. Since the entanglement of the connections is not altered, on average, it
follows that a lower threshold is found:
\begin{equation}
    \hat{E}_c=p_c^{\vartriangle} \approx 0.347.
\end{equation}
This proves that CEP is not optimal since it cannot generate long-distance
entanglement in the range $E(\varphi)\in (\hat{E}_c,E_c)$, whereas the quantum
entanglement percolation achieves it with a strictly positive probability.

It is interesting to note that this example has also been used
to show that the close analogy between quantum entanglement
and classical secret correlations can be applied to
entanglement percolation~\cite{LG11}. In this analogy, secret
key bits, rather than entanglement are shared between neighbouring nodes.

\subsubsection{Multi-partite entanglement percolation}
\label{percolation_pure_MEP}

We have seen that CEP can be enhanced by first
applying some quantum operations at the nodes~\cite{ACL07,PCA+08,LWL09}. All
examples proposed in these articles consist of transforming the quantum lattices
by a sequence of entanglement swappings, thus conserving the average entanglement
of the bonds. They are, however, restricted to purely geometrical transformations,
and they apply to specific lattices only. In this respect, it was not clear
whether the CEP strategy, and particularly the corresponding threshold,
could be improved in general. A positive answer to this question was
given in~\cite{PCL+10}, in which a class of percolation strategies exploiting
\textit{multi-partite entanglement} was introduced. To that end, one performs
the entanglement swappings in a more refined way, which we describe below.

\paragraph{Generalised entanglement swapping}
The key ingredient of the multi-partite method is to consider a
\textit{generalised entanglement swapping} at the nodes: First,
an incomplete measurement (i.e., not a complete
projection) is performed at the central node by applying the
operators
\begin{equation}
    \eqalign{
    M_1 &= \ketbra{0}{00} + \ketbra{1}{11},\\
    M_2 &= \ketbra{0}{01} + \ketbra{1}{10},}
    \label{eq:gen_ent_swap}
\end{equation}
for which the completeness relation $\sum_{i=1}^2
M_i^{\dagger}M_i^{\vphantom{\dagger}} = \id_4$ is
satisfied. This measurement leaves a two-dimensional subspace at the
central station  entangled with the two outer
nodes. Thus, the central station still plays a role in the
propagation of entanglement through the lattice. Second, not
only two but $n\geq2$ links are ``swapped'' at the same
time, and then the Procrustean method of entanglement
concentration is performed on the resulting state; see
\eref{eq:procrustean}. These operations succeed with a
finite probability that depends on the number of links and
on their entanglement~\cite{PCL+10}, generating the
Greenberger-Horne-Zeilinger (GHZ) state
\begin{equation}
    \ket{\GHZ_n} \equiv \frac{\ket{0}^{\otimes n}+\ket{1}^{\otimes n}}{\sqrt{2}},
    \label{eq:GHZn}
\end{equation}
which is the generalisation of the Bell pair $\ket{\Phi^+}$ to $n$ qubits.

\paragraph{Entanglement thresholds: from bond to site percolation}

\begin{figure}
    \centering
    \subfloat{\label{fig:MEP_a}
        \includegraphics[width=3.9cm]{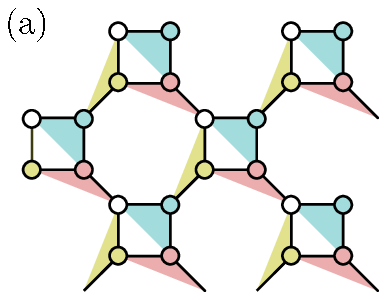}}
    \qquad\quad
    \subfloat{\label{fig:MEP_b}
        \includegraphics[width=3.9cm]{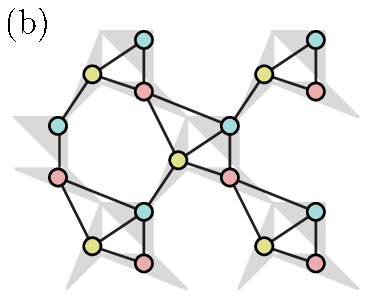}}
    \caption{Multi-Partite entanglement percolation.
    (a) Some nodes (filled circles) of the quantum lattice $\cL$
    probabilistically transform two links into a GHZ state on three nodes, which
    is depicted by a triangle. No operation is performed at the other nodes (empty circles).
    (b) The nodes of the new lattice $\hat{\cL}$ represent the GHZ states
    that have been created, and a bond is present whenever two GHZ states
    are adjacent in $\cL$.}
    \label{fig:MEP}
\end{figure}

In the multi-partite strategy one creates from the links $\ket{\varphi}$
of a quantum lattice $\cL$ a new lattice $\hat{\cL}$, where the
nodes represent the GHZ states obtained with probability $p$ by the generalised entanglement
swappings. Two vertices in $\hat{\cL}$ are connected by a bond if the
corresponding GHZ states share a common node in the original lattice.
This defines a site percolation process with occupation probability $p$,
and above the site percolation threshold of the new lattice, the
entanglement is propagated over a large distance as follows. Consider the
situation in which two GHZ states of size $n$ and $m$ sharing one node have been
created. One builds a larger GHZ state on $n+m-1$ particles with
unit probability by performing a generalised entanglement swapping on the two qubits
of the common node. This operation is iterated, eventually yielding
a giant GHZ state spanning the lattice. Then, given a GHZ state of any
size, a perfect Bell pair is created between any two of its qubits
by measuring all other qubits in the $X$ basis.

An example of multi-partite entanglement percolation is given in \fref{fig:MEP}.
In this case, since the probability to create a
GHZ state of three qubits is equal to that of creating a
Bell pair (only two links are required), the minimum amount
of entanglement of the bonds for generating long-distance
entanglement is $\hat{E}_c \approx 0.650$, which is equal to
the site percolation threshold in $\hat{\cL}$~\cite{NMW08}. Since the critical value for CEP is $E_c
\approx 0.677$ (the threshold for bond percolation in $\cL$~\cite{SZ99}),
 it follows that multi-partite entanglement percolation
surpasses CEP in the range $E(\varphi)\in(\hat{E}_c,E_c)$.\\

The previous example appeared in~\cite{PCL+10} together with many other
lattices for which the critical values using the multi-partite strategy are
lower than the thresholds for CEP. In particular, an improvement
is found for all \textit{Archimedean lattices}, which are
tilings of the plane by regular polygons (such as the square or the triangular
lattice). Moreover, it is shown that not only the thresholds but also the probabilities
$P(A\connected B)$ are better for any amount of entanglement $\hat{E}_c<E(\varphi)<1$.
This clearly indicates that the interplay of geometrical lattice transformations
and multi-partite entanglement manipulations is a key ingredient for propagating
entanglement in a quantum network.

\subsection{Towards noisy quantum networks}
\label{percolation_mixed}

We have already noted that creating perfectly entangled
states via LOCC by consuming a finite number of states on a
network of links of full-rank mixed states is not possible,
even considering perfect operations. Two directions we may
take from here are: i) If we cannot create a Bell pair, we may
try to create a state with the highest possible
fidelity. ii) Ask instead: For what class of mixed states it
indeed possible to create a Bell pair, and what are the
optimal protocols?

We have already paid some attention to the first question
above and will review a more detailed examination in
\sref{complex_complex_mixed}.  The answer to the second
question is based on two results: i) It is not possible to
transform a single copy of a mixed state to a pure state
with local operations, and ii) In the case of two-qubit
pairs, a pure state may only be obtained from two or more
pairs belonging to a certain class of rank-two
states~\cite{Jan02}, which we shall call purifyable mixed
states (PMS). Protocols on networks using this approach
were studied in~\cite{BDJ09,BDJ10}. This work was
extended to a hybrid approach addressing both the first and
second question in~\cite{BDJ10b}. The protocols
presented in these three articles are the subject of this
subsection. Note that what these authors call a PMS is a
slightly more restricted class of states.

The most obvious constraint on the design of entanglement
distribution protocols is that at least two disjoint paths
of PMSs must exist between two stations in order to have a
non-zero probability of a Bell pair between them. The final
stage must consist of purifying multiple PMSs.  We optimally
convert two PMSs to a Bell pair in two stages. First, we
perform a pure-state conversion measurement (PCM) as
follows.  We perform a quantum logic operation consisting of
unitaries called the controlled-NOT gate at each local
station, with qubits from one pair acting as targets in both
cases.  We then measure the targets in the computational
basis, and if both results are $1$,
we have generated a pure state. If the two input states were
identical, then the output state on success is already a
Bell pair. But, in general, another Bell pair conversion using the
Procrustean method according to \eref{eq:procrustean} in
\sref{concepts_entanglement_purification} must be
performed.  However, as we shall see below, it is sometimes
advantageous to delay this final Bell pair conversion and use
the intermediate state in a different way.

\paragraph{Swapping}
The behaviour of PMSs under swapping is similar to that of pure states.
As in the case of pure states, we project onto the Bell basis, but now only
two of the resulting states are useful, themselves being PMSs. Most importantly,
the fidelity of the average resulting state decays exponentially in the
number of links as it does for pure states.

\paragraph{CEP}
CEP protocols analogous to pure state protocols are defined
by taking regular lattices with multiple PMSs per
bond. This is the most basic way to provide the necessary
two pairs between nodes. This situation is similar to the
double-bond honeycomb lattice of
\sref{percolation_pure_QEP}, except that the states
are PMSs and there may be more than two pairs connecting
nodes.  We saw for pure states that a Bell pair conversion on
each bond succeeds with a probability $p$, mapping the
entanglement distribution problem to classical percolation
with bond density $p$. In the present case, we still map
directly to classical percolation, but the bond density $p$
is determined by the success rate of some conversion
protocol of the PMSs to a Bell pair.  For two PMSs, the
optimal protocol for this conversion is known~\cite{Jan02}
and has a maximum success rate $p=1/2$, so that, for
instance, percolation is possible on the double-bond
triangular lattice but not on the double-bond square
lattice. It is possible to achieve $p>1/2$ for three or more
bonds, but the optimal conversion protocol is not known in
this case. One protocol projects locally the entire state
onto the subspace that is pure and entangled~\cite{CLLH02}. 
This was investigated in~\cite{BDJ10},
along with better protocols that purify
multiple bonds in smaller groups and reuse states from some
of the failed conversions.

\paragraph{QEP}
We have seen that for pure-state entanglement percolation,
the first step in quantum pre-processing that goes beyond
CEP is to note that, for a chain of two pairs, swapping
before Bell pair conversion is better than Bell pair conversion
before swapping. For PMSs the simplest analogy has more
choices of when to do a pure-state conversion (PCM), Bell pair
conversion, or swap, because we need a minimum of two pairs
in parallel for each link in the chain. It turns out that
the optimal method for this case of two pairs per link is to
perform the PCM on each link and then to swap the resulting
pure states before doing a Bell pair conversion. If the input
states are identical, this is equivalent to CEP because, as
mentioned already, a successful PCM already returns
Bell pairs. In~\cite{BDJ10} this method was applied to
small configurations and the results, in turn, to some
regular and hierarchical lattices.

In all of these protocols, the only non-trivial case, with respect
to CEP, is when the multiple pairs making up a link are PMSs with differing parameters;
otherwise we get CEP again after the first purification step. In the case
where the parameters {\it are} different within a link, we must first do
a PCM, which maps the problem to the original pure-state percolation problem
with some of the links probabilistically deleted. A similar idea is used
in~\cite{BDJ10b}, where pairs in certain rank-three states replace the PMSs.
These can be converted to binary states via a sort of PCM that leave a
separable state on failure. The resulting lattice can be treated with error
correction methods as in \sref{correction}, with the difference that the pairs
in binary states are deleted probabilistically.

\subsection{Open problems}
\label{percolation_open}

It is natural to wonder about the optimality of the protocols based on
entanglement percolation, either for pure states or for certain classes of
mixed states. We focus on the square lattice here because
it is very common, but the arguments apply to other lattices equally well.

It is obvious that percolation protocols cannot be optimal
for every amount of entanglement of the links. In fact, we
have seen in \sref{percolation_det_regular} that a
long-distance perfect Bell pair can be obtained
deterministically if $E(\varphi)$ is larger than the
threshold $E_c\approx0.649$, whereas this is possible in
entanglement percolation only if $E(\varphi)=1$. On the
other hand, the purification method completely fails if
$E(\varphi)<E_c$, while CEP yields positive results in the
range $E(\varphi)\in(\frac{1}{2},E_c)$.  Consequently,
considering one strategy only for all values of entanglement
is not sufficient, but finding the optimum one for a given
value $E(\varphi)$ in the links is a formidably difficult
problem~\cite{PCA+08,LWL09,PCL+10}. In this respect,
multi-partite entanglement percolation is well-suited to
generate long-distance quantum correlations regardless of
the entanglement of the links, since it leads to high
connection probabilities and low thresholds at the same
time.  A somewhat more tractable question is:
\begin{description}
\item \noindent Does there exist a value of entanglement per link below which
  it is impossible to entangle two infinitely distant qubits
  using LOCC in a two-dimensional quantum network?
\end{description}
In the case of pure states the answer to this question is
not known~\cite{Per10b,PCL+10}. Strategies based on
multi-partite entanglement percolation are among the most
efficient ones that achieve this task (see chapter 2.4
in~\cite{Per10b} for a slight improvement of this scheme), but other efficient
protocols may exist.
For instance, it was shown in~\cite{PJS+08} that
long-distance quantum correlations can be obtained in a
square lattice using techniques of error correction (\sref{correction_example_GHZ}).

Quite surprisingly, the situation turns out to be opposite in the
case of mixed-state networks. In fact, suppose that the
connections of the network are given by the Werner state $\rhoW(p)
= p \ketbra{\Phi^+}{\Phi^+} +(1-p)\ \id_4 / 4$, with $p$ smaller
than the (classical) threshold $p_c$ for bond percolation in the
corresponding lattice. That is, the quantum state describing the
whole system is a classical mixture of lattices whose links are
either perfect Bell pairs or completely separable states. In the
limit of infinite size, however, \textit{none} of these lattices
possesses an infinite cluster of Bell pairs. The threshold $p_c$
is thus a lower bound on $p$ for a lattice of states $\rhoW(p)$
since, by definition, no local quantum operation can create
entanglement from separable states. In the square lattice, for
instance, genuine quantum correlations cannot be generated over
arbitrarily large distances if $p<p_c^{\Box} = 1/2$, even though
all connections are entangled in the range
$p\in(\frac{1}{3},\frac{1}{2})$. Finally, the situation for
mixed-state lattices in three dimensions is similar to that of
pure states: there exists a threshold to generate long-distance
entanglement in the cubic lattice
(\sref{constant_resources}), but no interesting lower
bound is known. In fact, the previous argument leads to a lower
bound given by the percolation threshold on the cubic lattice $p_c
\approx 0.2488$~\cite{LZ98}, but at this bound the quantum
connections are useless in any case since $\rhoW(p)$ is separable
for $p\leq\frac{1}{3}$. The previous argument can be
generalised to any mixed state using the concept of the {\em best
separable approximation} (BSA) to an entangled state, introduced
in~\cite{LS98}. Given a quantum state $\rho$, one decomposes it as
the mixture of an entangled and a separable state, $\rho_E$ and
$\rho_S$, with positive weights $p_S$ and $p_E$ such that
$p_S+p_E=1$:
\begin{equation}\label{eq:bsa}
    \rho=p_S\ \rho_S+p_E\ \rho_E .
\end{equation}
The BSA to the state $\rho$ is defined by the decomposition
maximising $p_S$. Clearly, if the states in a network are such
that the separable weight of its BSA $p_S$ is smaller than the
network percolation threshold, there is no protocol allowing
long-distance entanglement distribution.

In conclusion, it is of great interest to determine if there exists
a lower bound $E_{\min}$ for propagating the entanglement in quantum networks,
and, if this is the case, to design new
protocols that bring $E_c$ as close to $E_{\min}$ as possible.


%% file: sections/correction.tex
\newpage
\setcounter{footnote}{0}
\section{Network-based error-correction}
\markboth{\sc Network-based error-correction}{}
\label{correction}

\newcommand{\Lpsi}{\ket{\tilde\Psi}}
\newcommand{\Lzero}{\ket{\tilde 0}}

In the previous section, we showed that percolation allows one to
efficiently create entangled pairs of qubits over a large distance
in quantum networks that consist of pure states or of a restricted
class of mixed states. This is a considerable improvement over
one-dimensional systems, in which the probability to generate
remote entanglement decreases exponentially with the distance. The
connectivity of the nodes plays a key role in this respect, but it
is not clear if a similar effect exists in networks subject to
general noise, that is, described by mixed states of full-rank. In
fact, a finite number of such states cannot be purified into
maximally entangled states
(\sref{concepts_entanglement_purification}), which is an essential
requirement for entanglement percolation.

The aim of this section is to review the strategies that have been developed
for propagating the entanglement in quantum networks whose connections are full-rank
mixed states. However, we restrict our attention to Werner states
defined in \eref{eq:Wx}. This entails little loss of generality because
any mixed state can be transformed to a Werner state with the same fidelity
via depolarisation.

\subsection{A critical phenomenon in lattices}
\label{correction_introduction}

While most results on distribution of pure-state entanglement on lattices are based
on percolation theory, another critical phenomenon lies at the heart of the
propagation of mixed-state entanglement. Without being too rigorous,
let us describe here this phenomenon in the case of a (classical) square lattice;
the connection to quantum communication with mixed states is made
in the next sections.

\begin{figure}
    \centering
    \subfloat[][$p=0.02$]{\label{fig:nbec_a}
        \includegraphics[width=.28\linewidth]{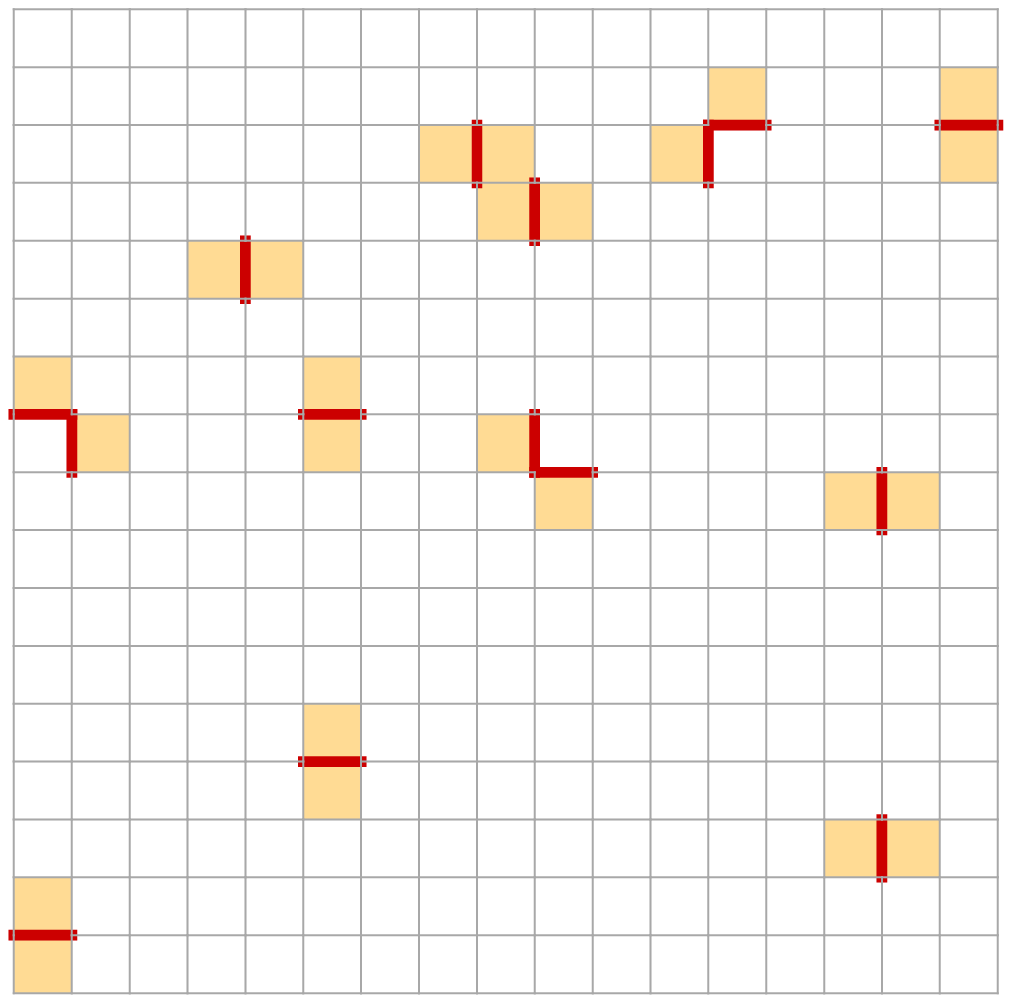}}
    \qquad
    \subfloat[][$p=p_c\approx0.11$]{\label{fig:nbec_b}
        \includegraphics[width=.28\linewidth]{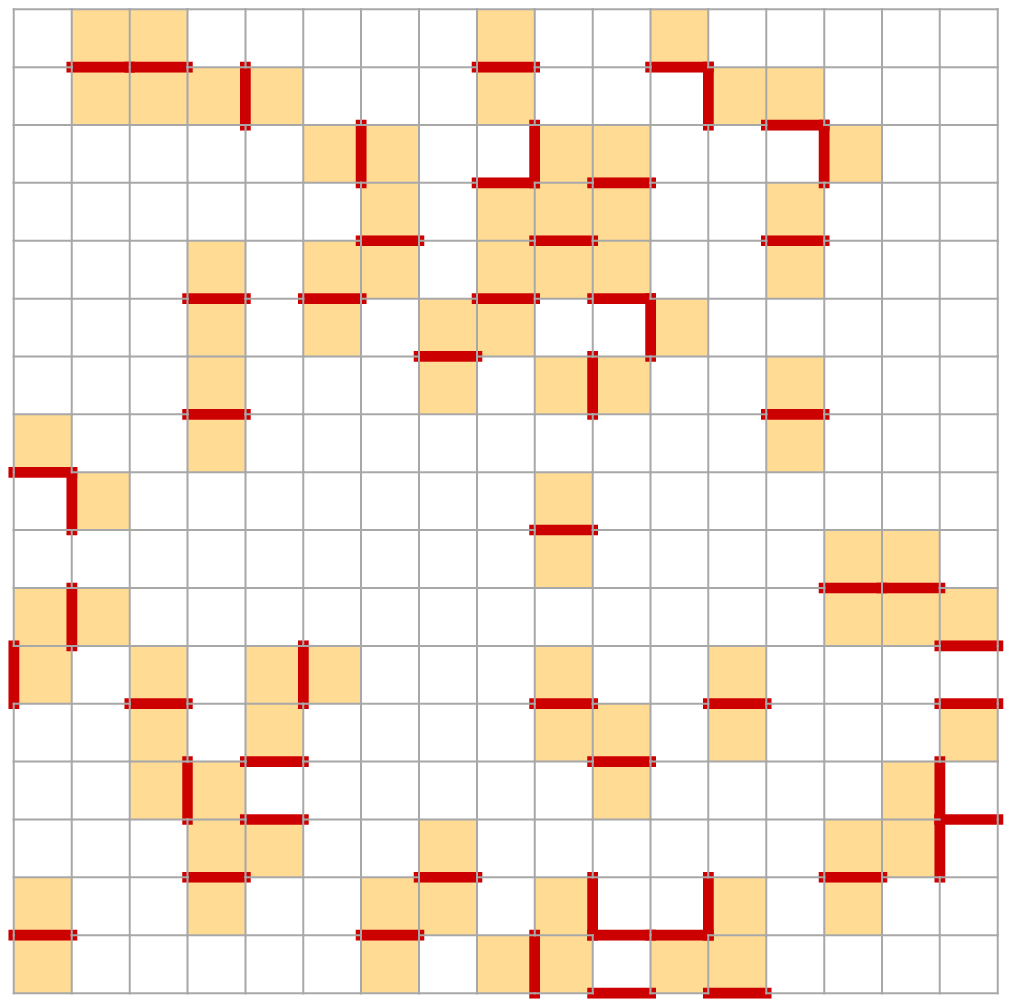}}
    \qquad
    \subfloat[][$p=0.25$]{\label{fig:nbec_c}
        \includegraphics[width=.28\linewidth]{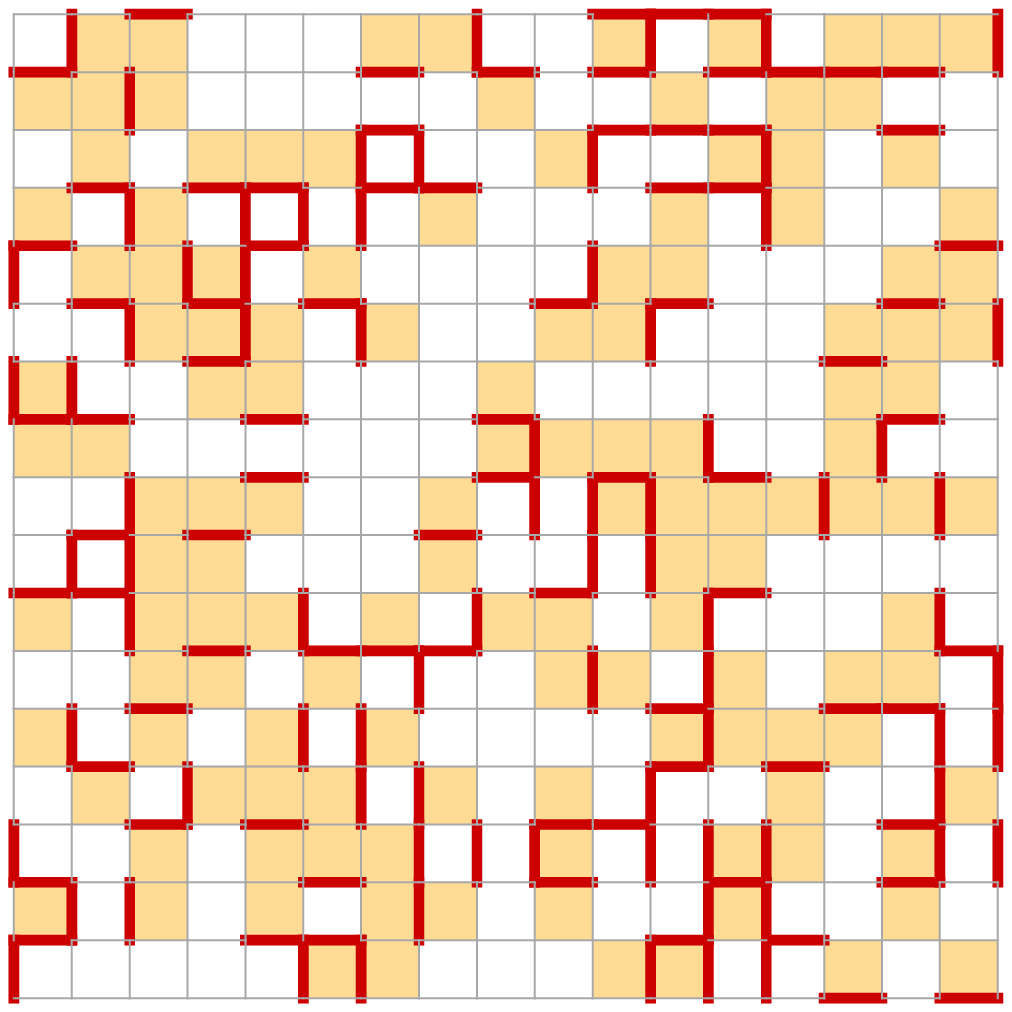}}
    \caption[Network-based error correction]{%
   In network-based error correction, syndromes (light orange squares)
   can be extracted from a lattice in which edges are defective 
   (heavy red lines) with probability $p$;
   the available information about the lattice concerns syndromes and
   not defective edges. For small $p$, one sees that syndromes come in
   pairs, so that most defective edges can be detected and thus corrected.
   Some detections may fail locally, for instance when paired syndromes are not
   adjacent, but this has no incidence at a macroscopic level. In an infinite
   square lattice, this remains true as long as 
   $p\lessapprox p_c$, whereas for
   larger values the error correction fails mostly everywhere.}
    \label{fig:nbec}
\end{figure}

Links of the lattice are randomly set to
``defective'' with probability $p$ and to ``valid'' with probability $1-p$,
but we suppose that one cannot test a link to determine its validity.
Instead, only a specific kind of information can be extracted from the lattice:
for each square, or more generally for each vertex of the dual lattice\footnote{
In graph theory, the dual $\Lambda^*$ of a lattice $\Lambda$ is defined as follows:
closed surfaces (polygons) in $\Lambda$ are mapped to vertices in $\Lambda^*$, and two
such vertices are connected if the corresponding polygons share an edge in $\Lambda$.
For example, the dual of the triangular lattice is the hexagonal lattice and
the square lattice is self dual.},
 we have access to the parity of adjacent links that are defective. Namely,
vertices of the dual lattice get the value 0 if the number of such links is even
and 1 otherwise. We call {\em syndromes}
those vertices which are set to 1, since they indicate that defective links lie
in their proximity. This fact is particularly obvious when $p$ is small, see
\fref{fig:nbec_a}.

Given a pattern of syndromes, one shows that most defective 
links can be detected if $p$ is smaller than a critical value
(\fref{fig:nbec} and \sref{correction_global_recovery}). In the remainder of
this section, we describe how this phenomenon can be used to create long-range
entanglement in mixed-state lattices.

\subsection{Correction of local errors from a global syndrome pattern}
\label{correction_global}

At present, few schemes have been proposed to generate long-distance
quantum correlations in noisy networks~\cite{PJS+08,FWH+10,Per10a,LCK12,GHH+12}.
Although the quantum operations that are performed at the nodes are quite
different for each scheme, the underlying principle is similar:
\begin{enumerate}
    \item The bonds are used to create a multi-partite entangled state that
        is shared by all nodes of the network. Due to the noise in the system,
        the generation of this state is imperfect;
    \item Local measurements on all but two distant qubits partially
        reveal at which places the noise altered the creation of the multi-partite
        state;
    \item A global analysis of the measurement outcomes determines the operations
        that have to be applied on the remaining two qubits in order to get
        useful remote entanglement.
\end{enumerate}
These steps are described in more detail in what follows, and the schemes
are reviewed in \sref{correction_example}.

\subsubsection{Creation of a multi-partite entangled state}
\label{correction_global_creation}
A first hint of the usefulness of multi-partite quantum states was
given in \sref{percolation_pure_MEP}. In that setting, a giant GHZ state is
created on the lattice by extracting perfect
entanglement from the bonds adjoining each node. Then, measurements in the $X$ basis
of all but two qubits transform the giant state into a Bell pair
between the two remaining qubits.
Finally, a local basis rotation depending on the measurement outcomes further
converts the Bell pair into, say, the maximally entangled state $\ket{\Phi^+}$.
The procedure is rather similar here, but the links of the networks are
used to create either a GHZ state (\sref{correction_example_GHZ}),
a surface code (\sref{correction_example_surface}),
or a cluster state (\sref{constant_resources}). The key point
of the construction is that while these states are simple enough
to be created by local operations on the nodes of the quantum network,
they also are tolerant of a certain amount noise, as described in the
following sections.

\subsubsection{Syndrome pattern}
\label{correction_global_pattern}
In the protocols involving pure states, it is known exactly whether a conversion of
partially entangled states into Bell pairs succeeds or fails, since this information
is given by the outcome of a measurement. In contrast,
in mixed-state quantum networks,
some noise enters the system randomly, and there is no way, \textit{a priori}, to know where this
happens. In fact, because every connection is a Werner state with non-unit
fidelity, the generation of the multi-partite state based on such connections
leads to a quantum state that contains errors (for instance, bit-flip and
phase errors on some of its qubits). Hence, if one decides to measure
all but the two target qubits right after the generation of the
multi-partite state, then the choice of the final basis rotation
will be correct with a probability of approximately only one fourth.
This means that we have no knowledge at all about which one of the
four Bell pairs we are dealing with, or in other words, the qubits are in a separable
quantum state.

One strength of the network-based error correction is that one can gain
some information about the errors without damaging the long-distance quantum correlations.
In fact, the multi-partite entangled states created in the quantum network
are highly symmetrical and satisfy a set of eigenvalue equations that can
be checked by local measurements: if the outcomes
do not match the symmetry of the target state at the corresponding nodes,
which is called a \textit{syndrome}, then one immediately knows that
at least one adjacent link inserted an error into the system;
see \sref{correction_example}.
The question of determining which link is responsible for the syndrome
is treated in what follows.

\subsubsection{Error recovery}
\label{correction_global_recovery}

\begin{figure}
    \centering
    \subfloat[]{\label{fig:syndromes_a}
        \includegraphics[width=.22\linewidth]{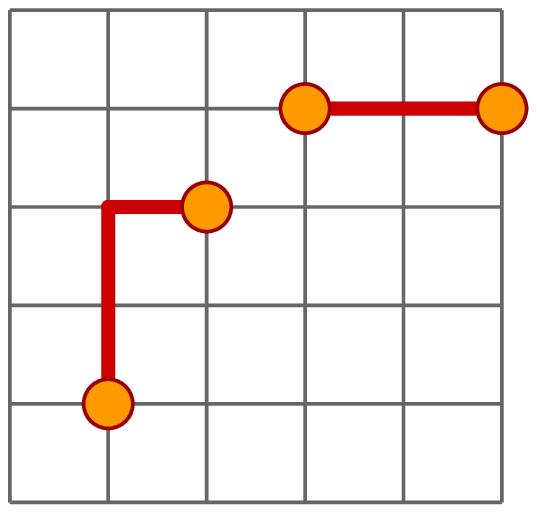}}
    \hspace{.05\linewidth}
    \subfloat[]{\label{fig:syndromes_b}
        \includegraphics[width=.22\linewidth]{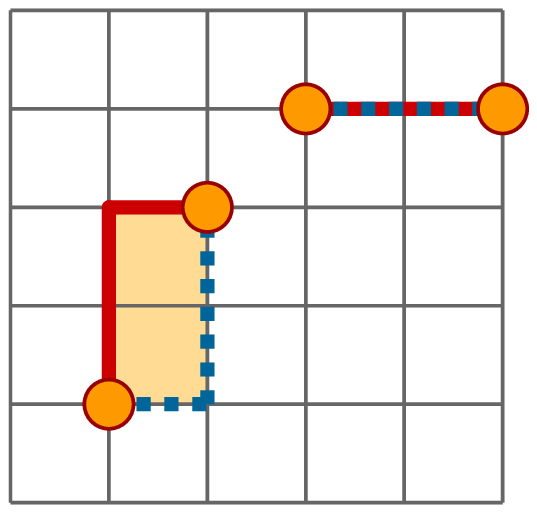}}
    \hspace{.05\linewidth}
    \subfloat[]{\label{fig:syndromes_c}
        \includegraphics[width=.22\linewidth]{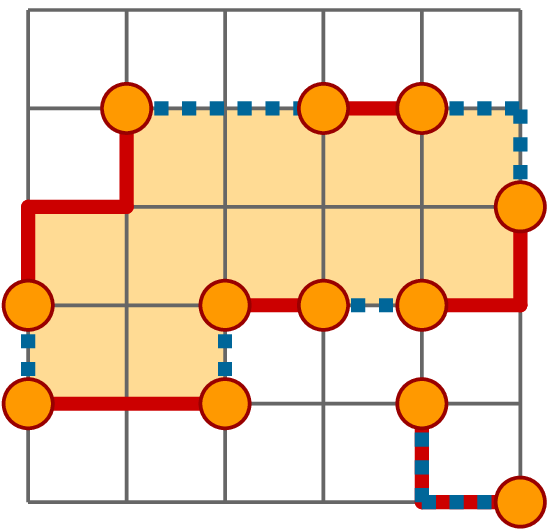}}
    \caption[Error correction from a pattern of syndromes]{%
        Error correction from a pattern of syndromes.
        (a) As the multi-partite entangled state is created
        from the noisy bonds of the quantum network (not shown), chains
        of errors (red lines) appear in the dual
        lattice (here, the square lattice), and syndromes
        (filled circles) are defined as the endpoints of the
        chains.
        (b) In dashed blue lines, a possible minimum-weight perfect
        matching of the syndromes. Inferring chains of errors
        that do not match the actual ones is equivalent to adding
        defective links, and closed loops of chains
        of errors are created.
        (c) The error correction fails if closed loops extend
        from one border to the opposite one. In a square lattice
        whose size tends to infinity,
        this happens if the error rate $p$ is larger
        than the critical value $p_c\approx 0.11$.}
    \label{fig:syndromes}
\end{figure}

Syndromes are defined on the nodes of the
dual lattice of the quantum network,
which is either a square lattice (first two schemes) or
a cubic lattice (third scheme).
The generation of the multi-partite entangled
state is such that the noise entering the system corrupts every link
of the dual lattice with probability $p$.
This creates \textit{chains of errors}, which are consecutive
corrupted links of the dual lattice. Syndromes correspond to the endpoints
of these chains and thus come in pairs, as depicted in \fref{fig:syndromes_a}.

If one knew the location of all chains of errors, then
it would be possible to perfectly restore the target multi-partite state.
The difficulty of the error recovery is that while the positions of the
syndromes are known, no other information about the chains is available. Since different chains
of errors can lead to a similar syndrome pattern, the recovery is ambiguous
and may lead to a wrong correction of the errors. However,
Dennis \textit{et al.} showed that,
in an infinite square lattice, a (partial) recovery is possible
if the error rate $p$ does not exceed a critical value $p_c$~\cite{DKLP02}.
This threshold is found via a mapping to the random-bond Ising model
and is approximately equal to $10.94\%$, which is
numerically calculated in~\cite{HPP01}.
In order to obtain long-distance quantum correlations
for $p<p_c$, one should be able to compute
all patterns of errors that lead to the measured syndromes
and then to choose the one that most likely occurred. This
is in-feasible in practice, but for small error rates $p$,
a good approximation of the optimal solution is the
pattern in which the total number of errors is a minimum.
In fact, such a pattern may be efficiently
found by a classical algorithm, known as the minimum-weight
perfect matching algorithm~\cite{Edm65B,CR99}. Illustrations of these ideas
are given in \fref{fig:syndromes_b} and \fref{fig:syndromes_c}.

\subsection{Examples of protocols}
\label{correction_example}
Now that the general concepts about network-based error correction have
been introduced, we describe in the following paragraphs the quantum operations
that are required to generate entanglement over long distance in noisy networks, as
proposed in~\cite{PJS+08,FWH+10,Per10a}.

\subsubsection{Independent bit-flip and phase errors}
\label{correction_example_GHZ}

Any entangled state of two qubits can be transformed by LOCC to the Werner
state defined in \eref{eq:Wx}, but to understand the scheme of~\cite{PJS+08}
it is more appropriate to consider a slightly different parameterization of a
two-qubit entangled mixed state:
\begin{equation} 
    \rho(\eps_b,\eps_p) \equiv \Big((1-\eps_b)(1-\eps_p),\,
    \eps_b(1-\eps_p),\,\eps_p(1-\eps_b),\,\eps_b\eps_p\Big)_{\cB}.
    \label{eq:rho-eps}
\end{equation}
In this equation, $\eps_b$ and $\eps_p$ stand for the probability that the
second qubit of the ideal connection $\ket{\Phi^+}$ has been affected by a
bit-flip and a phase error, respectively. This state is as general as a Werner
state in the sense that any quantum state of two qubits can be brought to this
form using LOCC only.%

\begin{figure}
    \begin{center}
       \includegraphics[width=.8\linewidth]{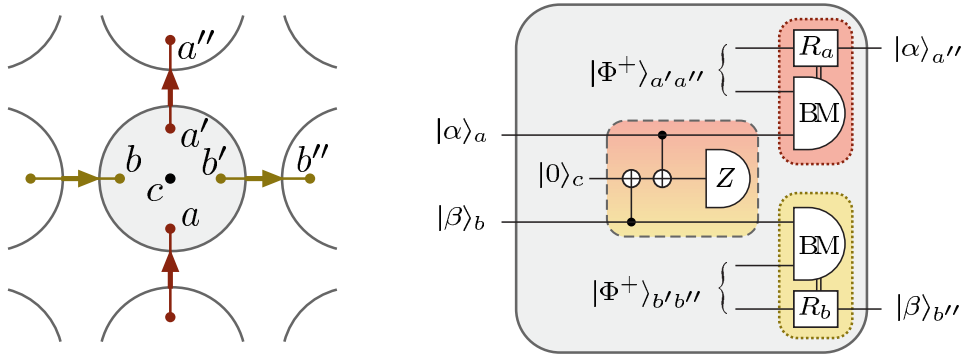}
        \caption[some text]{Generation of a giant GHZ state. Every station is connected to
        its neighbours by four entangled pairs of qubits. These links are consumed,
        via generalised Bell measurements followed by local rotations,
        to generate a GHZ state that spans the
        whole network. An auxiliary qubit ($c$) is used to check the parity of
        the ``incoming'' qubits $a$ and $b$: after two controlled-NOT
         gates\footnotemark
        it is measured in the computational basis. The station is then
        tagged with the outcome $0$ or $1$ of the measurement, which will
        be used to create the pattern of syndromes.}
    \label{fig:protocolGHZ}
   \end{center}
\end{figure}

\footnotetext{Under a controlled-NOT operation, the state
$\ket{11}$ becomes $\ket{10}$, $\ket{10}$ becomes $\ket{11}$,
while $\ket{00}$ and $\ket{01}$ are unchanged.}

\begin{figure}
    \centering
    \subfloat[]{\label{fig:bitflip_a}
        \includegraphics[width=.24\linewidth]{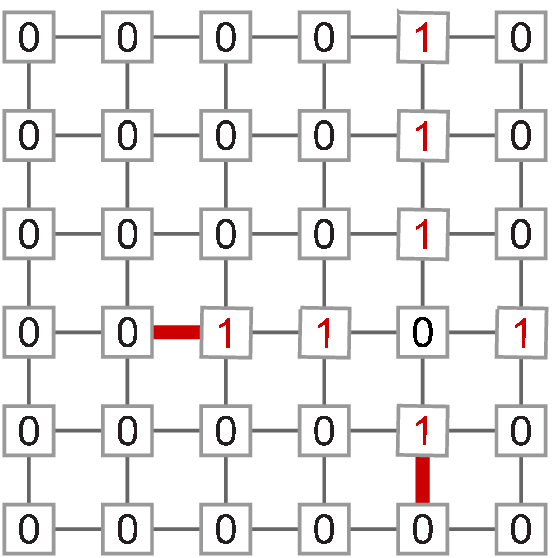}}
    \hspace{.05\linewidth}
    \subfloat[]{\label{fig:bitflip_b}
        \includegraphics[width=.24\linewidth]{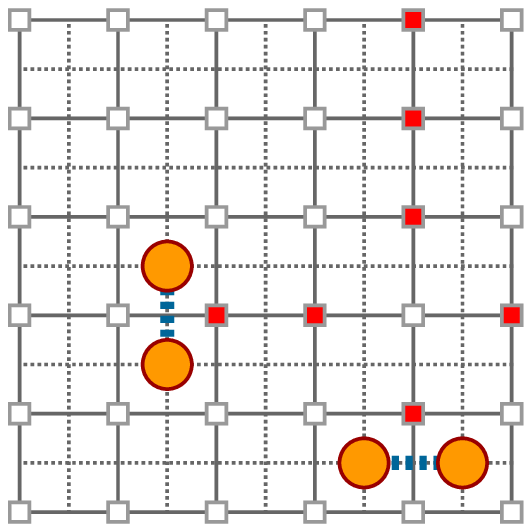}}
    \caption{%
        Syndrome pattern for the giant GHZ state.
        (a) Due to the sequence of measurements detailed in \fref{fig:protocolGHZ},
                 links (thick red lines) that insert a bit-flip error
        into the GHZ state invert the value of all rightward and upward parity checks.
        (b) Syndrome (circles) are created when an unit cell is surrounded
        by an odd number of parity check ``1'''s. The minimum weight
        perfect matching of these syndromes in the dual lattice (dashed
        lines) reveals the location of the noisy links.}
    \label{fig:protocolGHZ_syndrome}
\end{figure}

\paragraph{Protocol in the case of bit-flip errors only}
The links of a $N\times N$ square lattice are
used to create a giant GHZ state of $N^2$ qubits; see \fref{fig:protocolGHZ}.
For each direction, the stations perform the generalised entanglement
swapping described in \eref{eq:gen_ent_swap}. If one temporarily assumes that
phase errors are not present in the links of the network, then the resulting
multi-partite state is a mixture of GHZ states whose qubits are flipped with
probability $p=\eps_b$. If the error rate does not exceed the critical value
$p_c\approx10.9\%$, then most bit-flip errors can be corrected, as
depicted in \fref{fig:protocolGHZ_syndrome}.

The bit-flip error correction presented above can be applied to arbitrary
planar networks, as shown by Broadfoot \etal in~\cite{BDJ10b}. They also
prove that it can be generalised to entangled mixed states of rank three,
but another method has to be used in the case of full rank mixed states,
which is the topic of the following paragraph.

\paragraph{Protocol including both bit-flip and phase errors}
The global error correction works exactly as described above, but each qubit
is replaced by a logical qubit which is an encoded block of $n$ qubits.
Furthermore, all quantum operations on the logical qubit are implemented
by an appropriate protocol at the encoded level. Phase errors
are then suppressed by the redundancy of the following code:
\begin{equation}
    \eqalign{
        \ket{0} \mapsto &\ket{\tilde{0}}=\frac{1}{\sqrt{2}}
            \left(\ket{+}^{\otimes n}+\ket{-}^{\otimes n}\right),\\
        \ket{1} \mapsto &\ket{\tilde{1}}=\frac{1}{\sqrt{2}}
            \left(\ket{+}^{\otimes n}-\ket{-}^{\otimes n}\right),
    }
\end{equation}
where $\ket{\pm}$ are the eigenvectors of the $X$ basis.
Using majority vote, up to $t$ phase errors can be corrected
for each block of $n=2t+1$ qubits. A detailed discussion of
a fault-tolerant implementation of this encoding is done
in~\cite{PJS+08}, but here it is sufficient to state the final
result: long-distance quantum correlations can be obtained
with the encoded version of the global bit-flip correction,
but the number of links between neighbouring stations has
to increase logarithmically with the distance. The main difference with
the one-dimensional quantum repeaters is that all operations can be applied
simultaneously, so that no quantum memory is needed.

\subsubsection{Surface codes}
\label{correction_example_surface}

\begin{figure}
      \begin{center}
    \includegraphics[width=.45\linewidth]{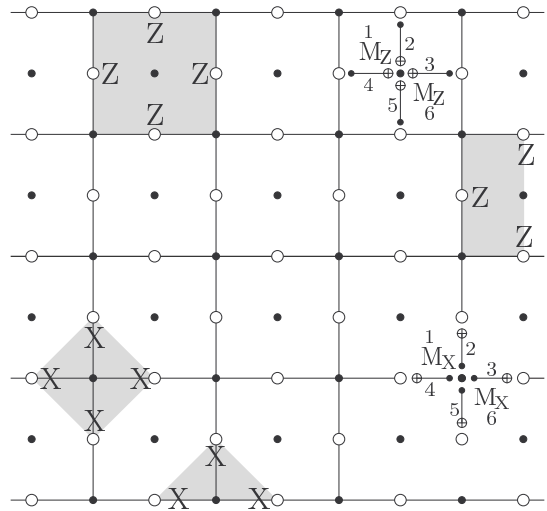}
    \caption{%
         Surface code of depth $d$  encoding a logical qubit. Data qubits are open circles.
    Syndrome qubits are filled circles.
   To guide the eye, four of the stabilisers are shaded grey.
   Two of the shaded stabilisers are on a boundary, so they have only three Pauli operators.
   The six steps in measuring a stabiliser are shown in detail: A preparation measurement,
   four CNOT gates, and final measurement.
   Adapted from A.G. Fowler {\it et. al.}, 2010. }
    \label{fig:surface_code}
      \end{center}
\end{figure}

Surface codes have been used to implement error correction in
quantum computation and communication~\cite{FWHL10,FWH11,HFDV11,SFH08,WFSH09,RH07}.
Here, we review three ways in which the surface code is used
to generate entanglement over large distances. The
protocol described in~\cite{FWH+10} differs from the
previous protocols in that the data to be transmitted is
encoded across the network as the multi-partite state is
created. In~\cite{LCK12}, the authors propose to use the
entanglement existing in the links of the network to apply
stabiliser measurements and to create the surface code
across the network. In both proposals the number of required
entangled pairs scales with the distance at which the
entanglement is to be generated. In what follows we describe
these two protocols, whereas the third one, which requires a
constant number of quantum connections only, is described
in \sref{constant_resources}.

\paragraph{Planar surface code}
For the planar surface code, half of the qubits encode the
data and half are only used to measure syndromes. The code
is implemented via stabilisers. A stabiliser of
$\ket{\psi}$ is an operator $A$ satisfying
$A\ket{\psi}=\ket{\psi}$.  Data is encoded by preparing the
array in a simultaneous eigenstate of a set of commuting
stabilisers, which are defined as follows. A lattice is
formed by associating an edge with each data qubit.  Each
face represents a stabiliser formed by the tensor product of
four Pauli $Z$ operators tensored with the
identity on all other data qubits. Vertex stabilisers are
defined similarly, substituting the $X$ operator for the
$Z$ operator. The exceptions are on the edges of the array,
where the stabilisers are products of only three Pauli
operators. The dimensions of the lattice are chosen to give
two boundaries with partial vertex stabilisers, and two with
partial face stabilisers, which ensures that the dimension
of the space satisfying the stabiliser conditions is $2$.
That is, the array encodes one logical qubit.
Details of surface codes and error correction
can be found in~\cite{DKLP02}.

\paragraph{Surface code communication protocol}
The planar surface code is implemented on a two-dimensional
rectangular array of entangled qubits and encodes a single logical
qubit; see \fref{fig:surface_code}.  We begin with a long rectangular
array that supports this code.  Square sections on the left and right
ends also each support a surface code. The left end is prepared in a
logical state $\Lpsi$ and the right end in the logical state $\Lzero$.
A sequence of operations is performed in such a manner that it spreads
the state $\Lpsi$ until the entire array encodes $\Lpsi$. Finally, we
perform some measurements that leave the array on the right end in the
state $\Lpsi$. Errors are measured during the entire procedure and
passed classically to the right end, where they are processed to
correct the result.

\paragraph{Implementing the surface code communication protocol}
The array is initialised with $\Lpsi$ on the left and
$\Lzero$ on the right, while all the data qubits in the
middle are measured in the $Z$ basis.  The state
$\Lpsi$ is then spread across the array by measuring each
stabiliser a number of times equal to code depth $d$,
which is the same as the height of the lattice. In general,
increasing $d$ allows better correction for gate errors.
The sequence of six gates shown
in \fref{fig:surface_code} is designed so that all
stabilisers may be measured simultaneously. As a result of
the repeated measurements, the syndromes
mark chains of errors that extend in time (for $d$ discrete
steps) as well as space. A change in a stabiliser
measurement signals a syndrome marking the beginning or end of a chain.
These error chains can be corrected in the same way
as described in the previous subsection.

\begin{figure}
      \begin{center}
    \includegraphics[width=.6\linewidth]{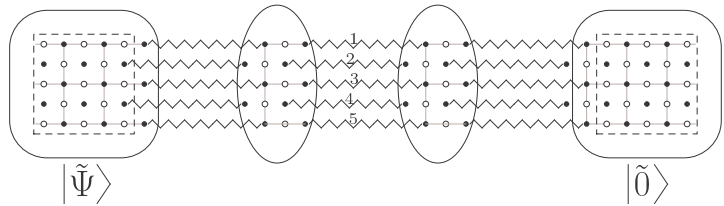}
    \caption{%
       Repeater based surface code. The middle section of
       the array of qubits is split at each syndrome qubit by
       coupling to a Bell pair.
      Adapted from A.G. Fowler {\it et. al.}, 2010. }
    \label{fig:repeater_surface}
      \end{center}
\end{figure}

As it stands, this protocol creates and transports entanglement only
through local interactions. Thus, it is not
sufficient for communication between distant parties.
For long distance communication, each syndrome qubit is replace by two
syndrome qubits, each of which is coupled to one party of a  Bell
pair as shown in \fref{fig:repeater_surface}.
The Bell pairs must be generated for each of the $d$ steps in
spreading the surface code.

Assuming no loss in transmission and that all gates within the
repeater nodes are perfect, the average time to failure of a
single link only grows non negligibly with the code depth $d$ if
the fidelity of the entangled pairs satisfies $F\gtrsim 0.92$. Furthermore,
the authors find that by modelling loss in transmission as
measurement in an unknown basis, loss rates
less than $0.45$ can be handled efficiently. They find that
realistic gate error rates do not affect these results
significantly. The number of qubits per repeater grows with
the code depth $d$, which scales as $\log(N/p_c)$, where $N$
is the number of links and $p_c$ is the desired
communication error rate. Reference~\cite{FWH+10} includes
a more thorough discussion of the error rate and
transmission rate.

\begin{figure}[tbp]\centering
\includegraphics[width =0.35\textwidth]{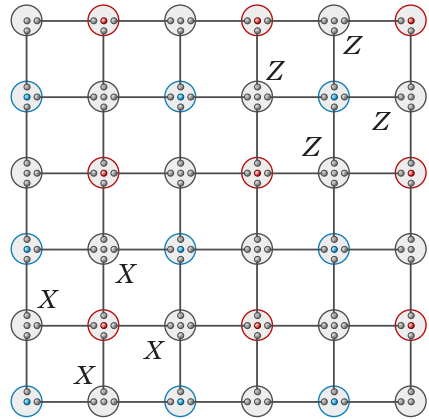}
\caption{
 Scheme to distribute entanglement in a 2D network based on
 the surface-error correction code. The nodes of the lattice
 are divided in black, red, and blue nodes. The entanglement
 shared between the nodes is used to measure stabiliser
 operators. Each red (blue) node performs $X$ ($Z$)
 measurements on its black neighbouring nodes according to
 the circuit described in \fref{circuitLi}. }
\label{schemeLi}
\end{figure}

\begin{figure}[tbp]\centering
\includegraphics[width = 0.6\textwidth]{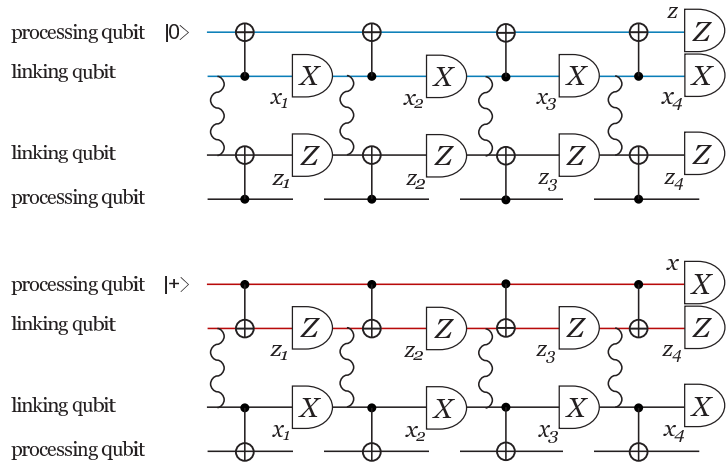}
\caption{Circuit to perform stabiliser  measurements using entanglement shared between neighbouring nodes. Each wave represents an entangled state. The measurement outcomes of the stabiliser measurement (e.g. $\bigotimes_{N_r} X_{N_r}$) is given by the product of the measurement outcomes at the neighbouring nodes (e.g. $x_1 \times x_2 \times x_3 \times x_4 $).}
\label{circuitLi}
\end{figure}

\paragraph{Network-based surface code  protocol}
The authors of~\cite{LCK12} suggest a different scheme to
generate long-distance entanglement based on the surface
code. The main idea is to use the entanglement present in
the links of the network to perform stabiliser measurements
on qubits sitting in different nodes. More specifically, the
authors consider a square lattice where, besides the four
qubits composing the network, each node has one extra qubit
(called a processing qubit); see \fref{schemeLi}. The nodes
are divided in three categories, black, red, and blue, in
such a way that each red or blue node is surrounded by four
black nodes. After initialising the processing qubits of
black nodes in the state $\ket{0}$, the entanglement shared
between neighbouring nodes is used to perform stabiliser
measurements. According to the circuit depicted
in \fref{circuitLi}, red nodes perform the stabiliser
measurements $\bigotimes_{\cN_r} X_{\cN_r}$, while blue
nodes measure $\bigotimes_{\cN_b} Z_{\cN_b}$ ($\cN_{r,b}$
denotes the neighbours of red or blue nodes).  If the
entanglement shared in the network is perfect, that is, if
the links are given by Bell states, the state of the black
processing qubits after the stabiliser measurements is
transformed into an eigenstate of the surface code. The last
step of the protocol consists in measuring all black
processing qubits except the ones that are held by Alice and
Bob.  The basis for these measurements is chosen to create a
maximally entangled state between Alice and Bob.

Due to imperfections, however, errors in the entanglement
shared in the network and in the operations result in
incorrect stabiliser-measurement outcomes. In order to
detect these errors, the stabiliser measurements are
repeated $N$ times. For each measurement run, a new
entangled pair of qubits must be generated between
neighbouring sites. Since the desired state is an eigenstate
of the stabiliser measurements, consecutive measurements
with differing outcomes indicate an error.  Finally, the
errors are corrected by pairing the error syndromes through
the network-based error correction.

The described protocol tolerates an error rate of
approximately $1.67\%$ in the quantum channels composing the
network, which, in turn, corresponds to links composed by
Werner states with $x\approx0.98$. One of the advantages of
the scheme presented in~\cite{LCK12} is that the same black
processing qubit can be used to measure its four
neighbouring qubits.

\subsubsection{Entanglement distribution with constant resources} 
\label{constant_resources}

In the error-correction based protocols described above, the
number of entangled pairs shared by neighbouring nodes has to
increase with the size of the network in order to generate
long-distance entanglement. On the contrary, it is shown
in~\cite{Per10a, GHH+12} that a constant number of
Werner states between neighbours is sufficient to achieve this task. In the
following, we discuss the proposal of~\cite{Per10a} which
uses cluster states in a 3D lattice. We then briefly
describe the main idea of~\cite{GHH+12}, which works both
for 2D and 3D lattices.

\paragraph{Three-dimensional cluster states}
The first protocol achieving long-distance entanglement with a
constant number of connections was proposed in~\cite{Per10a}.
In that article, it was shown that if the
fidelity of the connections is larger than a critical value
$F_c$, then entanglement can be generated between two qubits
$A$ and $B$ lying on opposite faces of a simple cubic
lattice of size $N^3$, with $N\rightarrow\infty$. In this
case, however, it must be stressed that the local quantum
operations are assumed to be perfect, which is not necessary
for the other strategies that are fault-tolerant.
In what follows, we describe in detail how this protocol works,
but let us first define the cluster state, which lies at the
heart of this proposal.

A cluster state $\ket{C}$ is an instance of graph states~\cite{HDR+06}, and it
is usually constructed by inserting a qubit $\ket{+}$ at each vertex of the graph
and by applying a controlled-phase%
\footnote{Under a controlled-phase operation, the state
$\ket{11}$ is multiplied by $-1$, while $\ket{10}$, $\ket{01}$,
 and $\ket{00}$ are unchanged.}
 between all neighbours. This state obeys the
eigenvalue equation $K_u \ket{C} = \ket{C}$ for all vertices $u$, where $K_u$
is the stabiliser
\begin{equation}
    K_u \equiv X_u \prod_{v\in \cN(u)} Z_v,
    \label{eq:correction_cubic_K}
\end{equation}
and where $\cN(u)$ stands for the neighbourhood of $u$ in the graph.
In~\cite{Per10a}, the desired controlled-phases are
non-local quantum operations, and therefore they have to be
performed indirectly through the use of the entangled
connections and by applying some generalised measurements at
the nodes. If one uses the noisy connections defined in \eref{eq:rho-eps},
some errors are introduced into the system and the cluster state is not
perfect anymore. Setting the bit-flip and phase error rates to $\varepsilon$, 
it can be shown that this results in local
$Z$ errors occurring independently at the nodes with a probability
$p\approx6\,\varepsilon$.

Three-dimensional cluster states are known to have an intrinsic capability of
error correction. In fact, long-range entanglement was shown to be possible between
two \textit{faces} of an infinite noisy cubic cluster state~\cite{RBH05}.
The difference between~\cite{RBH05} and~\cite{Per10a} is that local quantum operations
are assumed at every node in the latter case, in particular in the two faces.
The error correction runs as follows. First, all qubits but $A$ and $B$ are
measured in either the $X$ or the $Z$ basis. The measurement pattern is such
that, in the ideal case, the qubits $A$ and $B$ are maximally entangled:
\begin{equation}
    \eqalign{
    X_A X_B\ket{\psi_{AB}} &= \lambda_X\ket{\psi_{AB}},\\
    Z_A Z_B\ket{\psi_{AB}} &= \lambda_Z\ket{\psi_{AB}},}
    \label{eq:correction_cubic_XZ}
\end{equation}
where the eigenvalues $\lambda_X,\lambda_Z\in\{-1,+1\}$ depend on the measurement
outcomes $x$ and $z$ and are calculated from the stabiliser equations
\eref{eq:correction_cubic_K}. The effect of the local $Z$ errors
is to change the sign of these eigenvalues, thus ruining the quantum
correlations if no error recovery is performed. Then, the lattice is virtually
divided into two interlocked cubic sublattices (one for each
correlation $\lambda_X$ and $\lambda_Z$), and a parity syndrome is assigned to
nearly all vertices $u$:
\begin{equation}
    s(u) =
    \prod_{v\in\cN(u)} K_v =
    \prod_{v\in\cN(u)} X_v \prod_{w\in\cN'(u)} Z_w,
\end{equation}
where $\cN'$ designates the neighbourhood in the corresponding sublattice.
Since this equation arises from a product of stabilisers, we have that
$s=1$ if no noise is present in the system. However, $Z$ errors do not
commute with $X$ measurements, and the construction is such that an error
changes the sign of two syndromes that are neighbours in the other sublattice.

We are thus back to the error recovery described in \sref{correction_global_recovery},
with two differences nonetheless. First, of course, the lattice is three-dimensional
and not planar. Second, some syndromes lying on the opposite faces cannot
be assigned a value because of the specific measurement pattern, whereas
all syndromes are known in the usual error correction. This results in
imperfect quantum correlations, but the Monte Carlo simulations performed
in~\cite{Per10a} indicate that the distant qubits $A$ and $B$ are entangled
as long as the error rate $p$ is smaller than the threshold $p_c\approx2.3\%$.

\begin{figure}[tbp]\centering
\includegraphics[width = 0.6\textwidth]{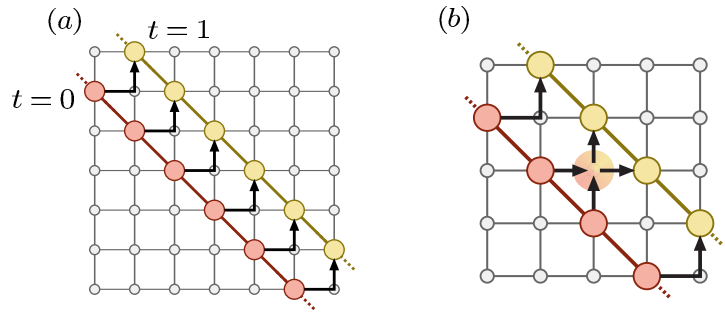}
\caption{
The task of distributing entanglement in a network can be
translated to a quantum-computation on a line with
next-neighbour gates. a) In a fixed time $t$, the qubits
composing the computer are given by a diagonal in the
lattice. Time evolution is translated in a teleportation
process between diagonals in the lattice. b) Two-qubit gates
between times $t$ and $t+1$ can be applied by teleporting
neighbour qubits to the same node in an intermediate
diagonal.}
\label{schemePers}
\end{figure}

\paragraph{Teleportation-based protocol}
In~\cite{PJS+08} it is shown that the problem of
transmitting entanglement in a two-dimensional network is
equivalent to the problem of fault-tolerant quantum
computation in a one-dimensional array of qubits restricted
to next-neighbour gates. Let us discuss how it works. First,
consider a square lattice in which we would like to create
long-distance entanglement, and let us suppose for the
moment that all links of the network are perfect. Consider
one of the diagonal array of qubits of this lattice as a
one-dimensional quantum computer at the initial time $t=0$;
see \fref{schemePers}a. By teleporting each qubit of the
computer twice, first right and then above, the state of the
computer is mapped to an upper-right diagonal array of
qubits. The computer is now at time $t=1$. If between $t=0$
to $t=1$ one has to implement a two-qubit gate between two
neighbour qubits, one first teleports one of the qubits up
and the other right; see \fref{schemePers}b. In this way,
they end up at the same location, where now the gate can be
locally applied. Finally, each qubit is further teleported
to the diagonal defining $t=1$.

In the case that each link corresponds to a Werner state,
the teleportation scheme can be seen as a quantum
computation where errors occur. In this way, a
fault-tolerant quantum computation scheme should be used. In
fact, such a scheme exists, where two qubits per site are
used~\cite{SFH08}.  Note that in fault-tolerant
error-correction schemes, one starts by encoding a known
state. However, in practice, in the very first stage the
qubits are exposed to noise and thus decohere. Furthermore,
in order to create an entangled state between two distant
nodes, one would need a decoding scheme that is also
fault-tolerant. 

In reference~\cite{GHH+12} the authors develop a fault-tolerant
encoding-decoding scheme into a 1D concatenated code~\cite{SFH08} and
into a 2D planar code~\cite{DKLP02} that works for unknown
states. Both the encoding and decoding protocols can be done in a
one-shot manner by measuring syndrome operators (products of $X$ and
$Z$ operators) and using error correction schemes based on syndrome
patterns. By combining the encoding-decoding scheme in 1D or 2D, with
the teleportation method shown in \fref{schemePers}, one can establish
long distance entanglement in the 2D and 3D square lattice network,
respectively.

A simpler scheme of entanglement distribution in a 3D lattice was also discussed
in~\cite{GHH+12}, where the authors use a 2D topological code.

\subsection{Conclusion}
\label{correction_open}

In this chapter, we have seen that
long-distance entanglement can be generated in mixed-state
networks. To this end, some information about where the noise
(bit-flip and phase errors) enters the system must be collected.
This is done by computing a series of parity checks at the nodes,
which creates a syndrome pattern. Then, the errors are corrected
by applying local unitaries that are determined by pairing, in an
optimal way, the detected syndromes. Note that all proposed protocols
are based on theoretical results for lattices of infinite size, so that it would be
interesting, if not necessary, to investigate their efficiency for
realistic networks of small or medium size.


%% file: sections/complex.tex
\newpage
\setcounter{footnote}{0}
\section{Networks with a complex structure}
\markboth{\sc Networks with a complex structure}{}
\label{complex}

\newcommand{\spath}{{\cal{P}}_{AB}}

Up to this point, we have seen protocols that generate
entanglement over a large distance in quantum networks with
a regular structure. In fact, as motivated in the
Introduction, the creation of remote entangled qubits is of
primary importance in quantum cryptography.
To date, real quantum networks have been designed in a
top-down, or executive fashion. But, as quantum information
technology progresses, we will eventually see networks in
which nodes and connections are added according to decisions
that are taken locally, rather than purely by executive design.
Such a process gives rise to self-organisation and complexity,
with the internet being the most relevant example.
In general, local organising principles give rise to
complex networks.
These networks describe a wide variety of systems in nature
and society modelling, among other
things, chemical reactions in a cell, the spreading of
diseases in populations.
We refer the reader to several books and reviews on this 
 complex networks~\cite{BS03,NBW06,AB02,New03,BLMCH06,DGM08}.
In the following, we first introduce the random graph model in
the context of complex networks. Then we review several studies
of distribution and concentration of entanglement on complex networks.

The simplest model that manifests some features of complex networks  was
introduced by Rapoport~\cite{Rap48,SR51}, and was treated in depth and rigorously
by Erd\H{o}s and R\'enyi~\cite{ER59,ER60,ER61}. In this model, known as
the \textit{random graph}, or \textit{Erd\H{o}s R\'enyi} (ER) model, each pair of
vertices in a graph is connected by an edge with probability $p$.

Although a great deal is known about the random graph, it lacks some of the
important features of real-life complex networks. However, it is
useful as a starting point, not only because it is easier to analyse,
but because it exhibits at least two of the most important features of
complex networks: 1) It possesses the small-world property, which
means that the length of the shortest path between two nodes increases
slowly with the system size. 2) It exhibits critical phenomena, with a
critical point and a single cluster with macroscopic density.  A
quantum version of the random graph was proposed in~\cite{PLAC10},
where it is shown that its properties change completely when they are
subject to the laws of quantum physics (\sref{complex_random}).

Many measures have been suggested to quantify the main
properties of real complex networks,
but three concepts seem to occupy a prominent place~\cite{AB02}:
the small-world~\cite{Koc89}, clustering~\cite{WS98}, and scale-free~\cite{BA99} behaviours. A graph
with the small world property typically has shortest paths
between nodes that scale like $\ln(N)$, where $N$ is the
number of nodes. Graphs with the small-world property
 may or may not have
properties such as community structure or some
regularity. But they have some links, perhaps a small
fraction, that connect random nodes, or distant nodes if
there is community structure. A graph with a large
clustering coefficient has the property that, if $A$ is
connected to both $B$ and $C$, then it is likely that $B$ is
also connected to $C$. A scale free graph has the property
that the probability that a node has $k$ links decays as a
power of $k$.  While the ER graph satisfies the small-world
property, it is not scale-free and has zero clustering
coefficient.  Many mathematical models have been introduced over
the years to include the two other characteristics. However,
in all of these networks, as in lattices, the method of
entanglement percolation is applicable and the percolation
thresholds are enhanced by some quantum strategies~\cite{CC09}; see \sref{complex_complex_percolation}.
In the following section, we treat yet another, but related,
critical phenomenon: the appearance of connected structures of a given shape in a random graph.

\subsection{Random graphs}
\label{complex_random}

\begin{figure}
    \centering
    \subfloat[][$p=0.10$]{\label{fig:crg_a}
        \includegraphics[width=.2\linewidth]{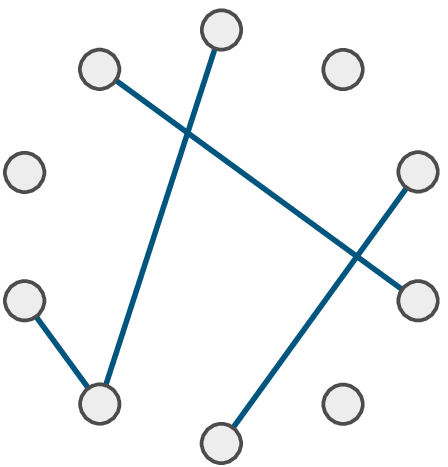}}
    \qquad\qquad
    \subfloat[][$p=0.25$]{\label{fig:crg_b}
        \includegraphics[width=.2\linewidth]{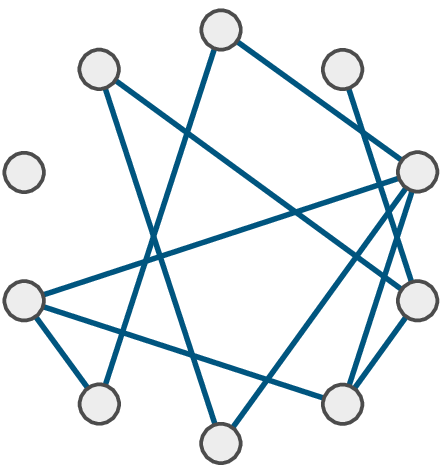}}
    \qquad\qquad
    \subfloat[][$p=0.50$]{\label{fig:crg_c}
        \includegraphics[width=.2\linewidth]{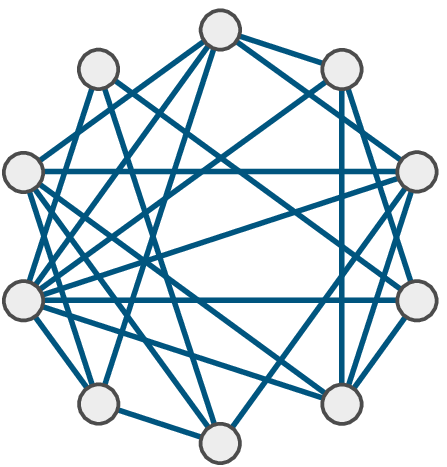}}
    \caption[Random graphs]{%
    Evolution process of a random graph of size $N=10$: starting
    from isolated nodes, we randomly add edges with increasing probability
    $p$, to eventually get a complete graph for $p=1$.}
    \label{fig:crg}
\end{figure}

Here, we briefly introduce random graph theory;
the interested reader is referred to~\cite{AB02} (and references
therein) for a more detailed description of these
graphs and a rigorous discussion of their properties.

The theory of random graphs considers graphs in which each pair of
nodes $i$ and $j$ is joined by a link with probability $p_{i,j}$.  In
the simplest and most studied model, the probability is independent of
the nodes with $p_{i,j}=p$. The generated graph is denoted $G_{N,p}$
and can considered to be the result of an evolution process: starting
from $N$ isolated nodes, random edges are successively added with
probability $p$ and the obtained graphs correspond to a larger and
larger connection probability; see \fref{fig:crg}.

\subsubsection{Appearance of subgraphs}
One of the main goals of random-graph theory is to determine the
probability $p$ at which a specific property of a graph $G_{N,p}$
typically arises, as $N$ tends to infinity. For fixed $p$, the typical
node-degree diverges with $N$, leading to a highly-connected, boring
graph. Instead, we let $p=p(N)$, with details of the dependence allowing
us to examine different interesting phenomena.

Many properties of interest appear suddenly, \ie, there
exists a critical probability $p_c(N)$ such that almost every
graph has this property if $p\geq p_c(N)$ and fails to have it
otherwise. Such graphs are said to be \textit{typical}. For instance,
it was shown that $G_{N,p}$ is fragmented into small isolated clusters
if $p \lesssim N^{-1}$, whereas percolation occurs above this threshold,
that is, one single giant component forms in the network.

A subgraph $F=(V,E)$ of $G_{N,p}$ is defined as a collection of $n\leq N$
vertices connected by $l$ edges. Like the giant cluster, the subgraphs
have distinct thresholds at which they typically form. It is proven
in~\cite{Bol85} that the critical probability $p_c$ for the emergence of $F$ is
\begin{equation}
    p_c(N)=c\,N^{-n/l},
    \label{eqn:complex_zcritic}
\end{equation}
where $c$ is independent of $N$. It is instructive to look at the
appearance of subgraphs assuming that $p(N)$ scales as $N^z$, with $z\in(-\infty,0]$
a tunable parameter: as $z$ increases, more and more complex subgraphs
appear; see \tref{tab:complex_thresholds}. In particular, only node-to-node
connections appear in the regime $z=-2$, whereas complete subgraphs
(of order four or more) emerge above the percolation threshold $z=-1$.

\newcommand{\crsub}[1]{\includegraphics[width=.6cm,bb=0 20 50 70]{#1}}
\begin{table}
    \begin{center}
    \begin{tabular}{c@{\quad}|@{\quad}c@{\qquad}c@{\qquad}c@{\qquad}c@{\qquad}c@{\qquad}c@{\quad}}
        $z$ & $-\infty$& $-2$ & $-\frac{3}{2}$ & $-\frac{4}{3}$ & $-1$ & $- \frac{2}{3}$\\[.2em]
        \hline\\[-1.5em]
         $F$ & \crsub{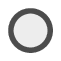} & \crsub{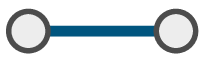} & \crsub{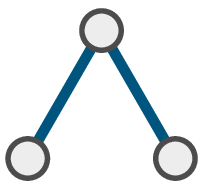} &
         \crsub{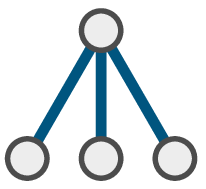} & \crsub{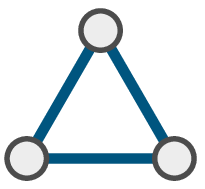}\ \ \crsub{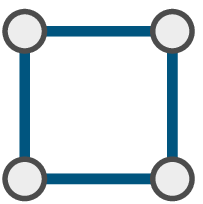} & \crsub{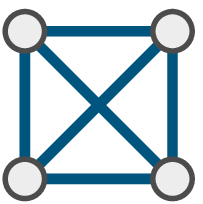}\\[.1em]
    \end{tabular}
    \caption[Thresholds for the appearance of classical subgraphs.]{%
    Some critical probabilities, according to \eref{eqn:complex_zcritic},
    at which a subgraph $F$ appears in random graphs of $N$ nodes connected
    with probability $p\sim N^z$. For instance, simple subgraphs appear at a
    small connection probability, whereas cycles and trees of all orders
    emerge at the critical value $z=-1$.
    After Albert and Barab\'asi~\cite{AB02}.}
    \label{tab:complex_thresholds}
    \end{center}
\end{table}

\subsection{Quantum random graphs}
\label{complex_random_quantum}
A natural extension of the previous scenario to a quantum context was considered
in~\cite{PLAC10}. For each pair of nodes, the probability $p_{i,j}$ is replaced
by a quantum state $\rho_{i,j}$ of two qubits, one at each node. Hence, every
node possesses $N-1$ qubits that are pairwise entangled with the qubits of the
other nodes. The study is restricted to the pure-state scenario and the pairs
of particles are identically connected, so that $\rho_{i,j}=\ket{\varphi}$.
A quantum random graph is then defined as
\begin{equation}
    \ket{G_{N,p}} \equiv \bigotimes_{i<j=1}^N \ket{\varphi}_{ij}.
\end{equation}
Expanding all terms of this expression in the computational basis,
one notes that this state is the coherent superposition of all possible
simple graphs on $N$ nodes, weighted by the number of states $\ket{11}$ they possess.
For instance, drawing a line for the state $\ket{11}$%
\footnote{Note that, contrary to the rest of the Review, a line denotes here a {\em separable}
 state rather than entangled qubits.}
 and nothing for $\ket{00}$, the quantum random graph on three nodes reads:
\newcommand{\crket}[1]{\includegraphics[width=.45cm,bb=0 16 50 66]{#1}\,}
\begin{eqnarray}
    \fl
    \ket{G_{3,p}} = \sqrt{\varphi_0}^{\,3} \Ket{\crket{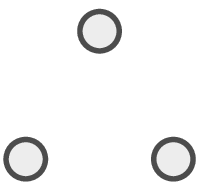}}
        +\varphi_0\,\sqrt{\varphi_1}\left(
        \Ket{\crket{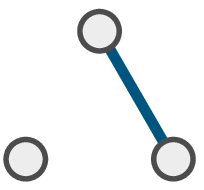}}+\Ket{\crket{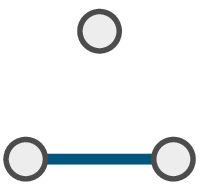}}+\Ket{\crket{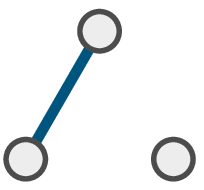}}\right)\nonumber\\
        +\sqrt{\varphi_0}\,\varphi_1\left(
        \Ket{\crket{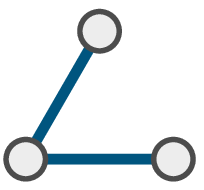}}+\Ket{\crket{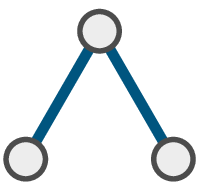}}+\Ket{\crket{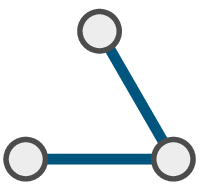}}\right)+
        \sqrt{\varphi_1}^{\,3}\Ket{\crket{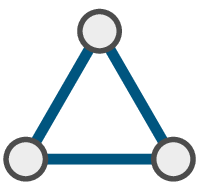}}.
\end{eqnarray}

The analogy with the ER model
is that one lets the degree of entanglement
of the connections scale with the number of nodes: $E(\varphi) = 2\varphi_1 \sim N^{z}$.
In this case, the ``classical'' strategy is to optimally convert each connection
of the graph, individually, into the Bell pair $\ket{\Phi^+}$, as described in
\sref{concepts_entanglement_purification}. The task of determining the type
of maximally entangled states remaining after these conversions is mapped
to the classical problem, and one obtains again the results of \tref{tab:complex_thresholds}
with $p = E(\varphi)$.

\subsubsection{A complete collapse of the critical exponents}
\label{complex_random_collapse}
The main result in~\cite{PLAC10} states that, in an infinitely large quantum random
graph and in the regime $p\sim N^{-2}$, the state
\begin{equation}
    \ket{F} \equiv \bigotimes_{i=1}^l \ket{\Phi^+}_{E_i},
    \label{eq:complex_F}
\end{equation}
which corresponds to \textit{any} subgraph $F=(V,E)$ composed of $n$ vertices
and $l$ edges, can be generated with a strictly positive probability using
LOCC only. All critical exponents of \tref{tab:complex_thresholds} thus
collapse onto the smallest non-trivial value $z=-2$. This clearly indicates
that the properties of disordered graphs
change completely when they are governed by the laws of quantum physics. It
is not the purpose of this Review to prove this result, but let us describe
the main quantum operation that lies behind its proof. In fact, it is another
example of a generalised measurement on many qubits that enhances the
distribution of entanglement in quantum networks.

\paragraph{A joint measurement at the nodes}
As for multipartite entanglement percolation
(\sref{percolation_pure_MEP}), allowing strategies that entangle
the qubits within the nodes yields better results than a simple
conversion of the links into Bell pairs. In the current case, the
construction of a quantum subgraph $\ket{F}$ is based on
an incomplete measurement of all qubits at each node.
More precisely, the measurement operators are projectors onto the
subspaces that consist of exactly $m$ states $\ket{1}$ out of
$M=N-1$ qubits:
\begin{equation}
    P_m \equiv \sum_{\pi_m} \pi_m^{\vphantom{\dagger}}
        \ketbra{\underset{M-m}{\underbrace{0\ldots0}}\underset{m}{\underbrace{1\ldots1}}}{0\ldots 01\ldots 1}
        \pi_m^{\dagger},
\end{equation}
where $\pi_m$ denotes a permutation of the qubits. Applied on all nodes of
the quantum random graph, this measurement generates a highly entangled
state shared by a random subset of nodes. The quantum correlations
corresponding to $\ket{F}$ are then extracted by a
suitable series of local operations at these nodes~\cite{PLAC10}.
Note that this measurement allows not only the creation of any quantum subgraph,
but also the generation of some important multipartite states,
such as the GHZ states~\cite{Per10b}.

\subsection{Percolation}
\label{complex_complex_percolation}
One may ask what pure-state entanglement percolation looks
like on complex networks rather than regular lattices. It is
obvious that CEP on complex networks works exactly as it
does on regular lattices; but it is not obvious how to
design pre-processing for QEP. However, a particular QEP protocol
has been applied with success to a variety of complex networks with
double-bond, pure, partially-entangled states~\cite{CC09,CC11}. As in the many cases of
links with multiple pairs that we have seen, this setup
makes possible the systematic application of a simple local
transformation, with the hope that general global effects
can be understood.

\begin{figure}
    \centering
    \includegraphics[width=.6\linewidth]{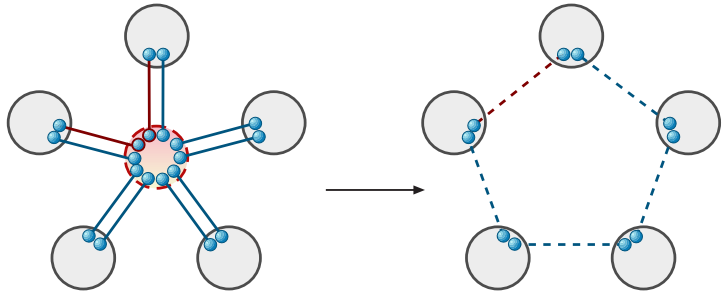}
        \caption{$q$-swap protocol. The red dots and lines
         are involved in one of five swaps that transform
         star into the cycle. Solid lines are partially entangled
         pure states. Dotted lines are post-swapping states.
         }
    \label{fig:qswap}
\end{figure}

Following~\cite{CC11}, we consider here networks where
each link between nodes consists of two pairs in the state
$\ket{\varphi}$; see~\eref{eq:phi}. CEP consists of the
optimal conversion of each pair of pairs to a Bell pair,
followed by swapping. The QEP protocol consists of
optionally performing $q/2$ swaps at a node of degree $q$,
with $q$ even. The swaps join pairs of links terminating at
the node. Thus, the node is replaced by a cycle whose bonds
are occupied with states according to the random outcome of
the swapping; see \fref{fig:qswap}. We call the original node and its bonds a
$q$-star, the transformed object a $q$-cycle, and the
transformation a $q$-swap.  Whether it is advantageous to
apply the $q$-swap depends on the details of the network.

The interesting parameter regime for complex networks is
often the tree-like regime-- that is, when there are so few
links that the probability that there is more than one path
connecting two nodes is negligible.  For instance, the
percolation transition of the ER network is in the tree-like
regime. The usual way to analyse tree-like networks is with
generating functions and recursion, which gives exact
results in this limit.  Although the $q$-swap creates loops,
they are small and isolated, so that an entire $q$-cycle can
be treated at one step in the recursion.  Still, one has a
choice in applying $q$-swap because, for instance, it cannot
be applied at neighbouring nodes. In~\cite{CC11} this
calculation was done for a breadth-first application, that
is, applying the $q$-swap at neighbour of a starting node,
and then at next-nearest neighbour's, where possible, etc.
Exact results were obtained for various networks.  For
instance, it was shown that any application of $q$-swap on a
tree lowers the threshold with respect to
CEP; see \sref{percolation_det_hierarchy}. The
generating-function analysis shows that, in the tree-like
regime, the optimal strategy with node-degree $k$ is to
either always or never perform $q$-swap depending on the
lattice and $k$. For other lattices this fact comes into
play.  For example, on the ER network, the optimal
application of $q$-swap yields the best results for average
degree near $4$ with a 20\% reduction in the threshold, with
similar results for the un-correlated scale-free network.

The presence of small cycles is recognised as a key feature of
many complex networks. In a network of scientific collaborators, for instance,
one can identify small groups, each member of which has been a coauthor with
each of the others. Of course, this phenomenon cannot be modelled in the
strict tree-like limit. Thus other models and
techniques of analysis, largely numeric, are required.
For instance, the Watts-Strogatz model begins with a one-dimensional chain
of nodes connected with nearest-neighbour's and
next-nearest-neighbour connections.  The terminal node of
each bond is then randomly rewired with small probability
$p$. The original local connections give the model community
structure, while the rewired bonds provide a few long-range
links that drastically reduce the shortest path between
nodes. Numerical simulations showed less impressive improvements than
the tree-like models. Still, the critical threshold is lowered by
$q$-swapping. However, this form of QEP is not always
advantageous. For large enough initial entanglement per
link, the CEP produces a Bell pair with probability $1$ and
the giant connected component is larger for CEP than for
$q$-swap.

\subsection{Mixed state distribution}
\label{complex_complex_mixed}
\begin{figure}
    \centering
    \includegraphics[width=10.9cm]{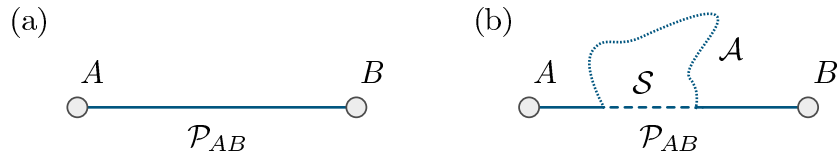}
    \caption{Establishing entanglement between nodes $A$ and $B$.
  (a) The shortest path $\spath$ between $A$ and $B$; the
geometry of the path is irrelevant, so we represent it by a straight
 line with individual links not shown. Other paths connecting $A$ and
$B$ are not shown.
  (b) The shortest path $\spath$ (solid line with a dashed segment)
  between $A$ and $B$.
Between the endpoints of  subpath ${\cal S}$ (dashed segment) there is an alternate path
${\cal A}$ (dotted line).}
    \label{fig:subpath}
\end{figure}

Here we discuss distributing full-rank mixed states on
complex networks.  One approach is to admit that the
exponential decay of entanglement due to swapping imposes
an upper limit on the number swaps that may be performed, before
all entanglement is lost.
The distance corresponding to this limit can be included in calculations of statistical
properties of entanglement. For instance, if this limit is
smaller than the correlation length, then the network is
essentially fractured with respect to entanglement via
direct swapping. This approach was taken numerically and
with generating functions in~\cite{CC11}. On the other hand,
it is {\it a priori} possible to approximate maximally
entangled links between distant nodes by concentrating links
from ever larger numbers of paths. Studies to date have
examined possible building blocks to this end~\cite{LPLA11}.
This work quantifies the gain in concurrence obtained
between nodes by employing concentration and distribution
protocols on multiple paths.  The most detailed results were
obtained for the simplest case of the single purification
protocol (SPP), in which the shortest path $\spath$ between $A$ and $B$ is
identified, and then a shortest path between two nodes on $\spath$ is
found. Both paths can be used to achieve a final entanglement
between $A$ and $B$ that is larger than that achieved by using $\spath$ only;
see \fref{fig:subpath}.

In particular, we examine here the results of the SPP protocol
applied to the ER random graph~\cite{LPLA11}.  The
initial network is the same as in the previous section,
except that the links are Werner states \eref{eq:Wx} rather
than pure states. As mentioned above, it is impossible
to extract a pure state from a finite number
of Werner states. We instead search for a protocol that
produces the mostly highly entangled mixed states
possible. More specifically, we seek to maximise the average
of the concurrence \eref{eq:Cmixed},
\newcommand{\avcon}[1]{\bar{C}_{#1}(x)}
\begin{equation}\label{defavcon}
 \avcon{}=\frac{2}{N(N-1)}\sum_{\alpha,\beta}\pi_{\alpha,\beta}C(\alpha,\beta;x),
\end{equation}
where the sum is over all pairs of nodes in the network, $\pi$ is the probability
that the protocol connecting $\alpha$ and $\beta$ was successful and 
$C(\alpha,\beta;x)$ is the resulting concurrence between the pair.

The most naive method to entangle two nodes $A$ and $B$ is
to perform entanglement swapping repeatedly between Werner
states along the shortest path $\spath$ joining $A$ and $B$
according to~\eref{eq:Werner_swap_C}; see
\fref{fig:subpath}a. However, in some cases, higher
entanglement may be obtained by additionally concentrating
the entanglement from an alternate path connecting
intermediate nodes on the path $\spath$ as shown
in \fref{fig:subpath}b.  In this single-purification protocol,
 one swaps along the
subpath ${\cal S}$ as well as the shortest available
alternate path ${\cal A}$, then performs a purification on
the two resulting parallel states, and finally performs
swapping at all remaining links. Whether a higher
entanglement results on average than that from simply
swapping along $\spath$ depends on the lengths of the paths
and the Werner parameter $x$.

\begin{figure}
        \centering
    \includegraphics[width=.5\linewidth]{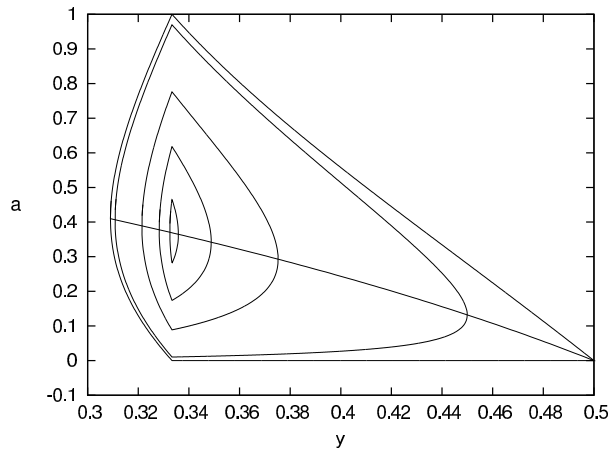}
     \caption{Regions in which the single purification
 protocol is advantageous {\it v.s.} rescaled Werner
 parameter $y$ and path length $a$, and for various values of excess
 alternate path length $b$.  The protocol yields higher
 average concurrence inside the closed curves. From the
 outermost to innermost curve the values of $b$ are
 $0,0.01,0.07,0.11,0.135$.  The curve cutting through the
 closed curves is $y^a=2y$ and maximises (independently of
 $b$) the increase in average concurrence with respect to
 $a$.
    \label{fig:aroots}}
\end{figure}

For instance, consider the parameters $L=||\spath||$, $a=||{\cal S}||/L$,
$b=(||{\cal A}||-||{\cal S}||)/L$, and $y=x^{1/L}$.  A
detailed analysis of the region in this parameter space for
which this purification protocol yields a higher average
concurrence is given in~\cite{LPLA11}, and is shown
graphically in~\fref{fig:aroots}.

\begin{figure}
        \centering
    \includegraphics[width=.5\linewidth]{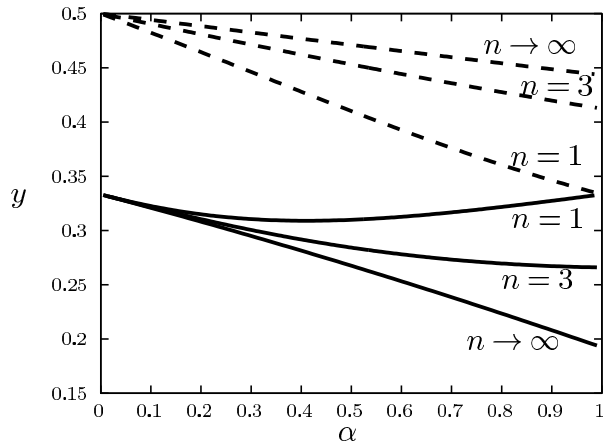}
     \caption{Regions in which a serial multiple purification
 protocol is advantageous {\it v.s.} rescaled Werner
 parameter $y$ and total fractional path length $\alpha$,
 and for various numbers of subpaths $n$. The multiple path protocol is advantageous only
 between the solid and dotted lines.
 \label{fig:multiple_pur}}
\end{figure}

Of course a network may offer more possibilities than
concentrating the entanglement from a single neighbouring path.
One could repeatedly concentrate entanglement from paths in
parallel or series. Let us consider the latter case, where instead
of a pair of paths ${\cal S}$ and ${\cal A}$, we have $n$ pairs
and that together the $n$ paths ${\cal S}_i$ cover a fraction
$\alpha$ of $\spath$. For simplicity we take $||{\cal S}||=||{\cal
A}||$. The region in parameter space for which this protocol is
advantageous were computed exactly and are shown
in~\fref{fig:multiple_pur}. One sees that many short paths are
better than few long paths, but that the maximum extent of the
good region approaches a limit as $n\to\infty$.

Suppose we apply only the single-purification protocol at
every possibility on a network.  Computing the average
concurrence over a random network is in general
difficult. However, the ER network at its critical point
$Np=1$ is in the tree-like regime and has further
statistical properties that facilitate calculations.
In particular, the exact asymptotic increase in concurrence over the naive
swapping protocol as a function of  Werner parameter $x$ and number of
nodes $N$ was found to be $\Delta\bar{C} \sim
AN^{-2} (1-x)^{-4}$.  Here $A\approx 6.5 \times 10^{-5}$ is
a constant that is easily computed by numerical integration. Although
this expression diverges as $x\to 1$, this occurs outside
the asymptotic regime where the purified paths have lengths
much shorter than the radius of the giant cluster.

It is important to note here that the above protocol is not capable
of entangling arbitrarily distant nodes independently of the initial
concurrence per link, but it is one that can be applied in principle
to arbitrary networks.

The preceding protocols, although treating mixed states, have assumed
perfect operations. If we further assume the noise model described in \eref{eq:noisy_gate}
and \eref{eq:noisy_measurement} we find that the protocol is rather sensitive to
noise. The results are quantitatively similar to~\cite{DBCZ99}
with the advantage destroyed for noise levels larger than a few percent.
For instance, the maximum average concurrence gain from the single
purification protocol $\Delta C$ was found to be
\begin{equation}
 \Delta C = \frac{1}{4} \Bigg\{\frac{4(1-\delta)^2}{9(1-2\delta)}
          -\frac{1}{3}(1+2\delta)-\alpha  \Bigg\},
\end{equation}
where $\delta=2\eta(1-\eta)$ and $\alpha=1/p_2^2-1$.

The richer variety and topologies of complex networks provide a fertile terrain
for developing more sophisticated distribution protocols. The studies reviewed
here make only the first steps. In particular, they apply relatively simple
quantum procedures for which averages over quantum outcomes and classical disorder
may be performed. Other important directions remain completely unexplored,
such as the dynamic creation and distribution of entanglement.

\subsection{Optimal path for distributing entanglement on networks}

Here we consider mapping entanglement distribution problems to
classical graph algorithms.  The majority of efficient classical
algorithms for solving various shortest-path problems on graphs assume
that the measure of path length is the sum of edge weights.  This
includes measures given by the product of non-negative edge weights,
which are mapped to this additive class by considering the logarithm
of the weights.  A few of these shortest-path problems are: 1)
shortest path between two nodes; 2) shortest paths between one node
and all other nodes; 3) problem 1) or 2) restricted to positive edge
weights.  The key reason that many problems using these measures admit
efficient algorithms is that they possess \textit{optimal
substructure} (also referred to as Bellman's \textit{optimality principle}):
 In this context, if $C$ is a node on the shortest path
from $A$ to $B$, then the later is composed of the shortest path from
$A$ to $C$ and the shortest path from $C$ to $B$.  These topics have
been researched and applied intensively and broadly for several
decades. References~\cite{CLRS09} and ~\cite{JUN07} are
two of the most popular textbooks treating the subject.

It is therefore of interest to know which entanglement distribution
problems can be expressed as a classical shortest-path problem. For
instance, suppose that each link $i$ is initially in a Werner state
$\rhoW(x_i)$, with the parameter depending on $i$.  Then,
using \eref{eq:Werner_swap_C}, we see that the
entanglement obtained by performing swapping at all nodes along a path
becomes a path length if the weights are taken to be
$-\log(x_i)$. Then, the path that yields the maximum entanglement
between nodes $A$ and $B$ via a series of swaps is the shortest path
between the nodes.

Another example is the question: Given the task of performing a series
of teleportations on a network to transfer a state from node $A$ to
node $B$, may we efficiently choose the optimal path?  In
reference~\cite{DiF12}, the question is addressed using the following
model.  Assume the state to be transferred has the 
form \eref{one_qubit}, with the restriction $\theta=0$;
That is $\ket{\phi}= \sqrt{\alpha_0}\ket{0} + \sqrt{\alpha_1}\ket{1}$\footnote{%
In this case the restriction $\theta=0$ is significant, as
we shall consider an average over these states.}.
The links are initially the partially entangled pure states described
by~\eref{eq:phi}, but with Schmidt coefficients
$\phi_{0,i},\phi_{1,i}$ depending on the link index $i$.
The protocol consists of teleporting a state $\ket{\psi}$
from $A$ to a neighbouring node, and then teleporting from this node
to yet another, and so on, until the state arrives at $B$.
Before continuing, we make a brief digression into classification of
protocols. All protocols involving pure states that we reviewed thus
far are so-called \textit{probabilistic} protocols; they involve
recording a measurement and thus retaining information about the
resulting state, which is one of a number of possible states, each
obtained with a certain probability. Another kind of protocol, called
a \textit{deterministic} protocol, is used in the present scenario.
In this case we don't record the measurement, so that the output state
is a mixture of a number of possible outcomes. Thus, even with a pure
initial state and pure link states, and perfect operations, this
results with probability one (\textit{i.e} deterministically) in a
mixed state $\rho$.  A measure of how close the state received at node
$B$ is to the initial state at node $A$ is the fidelity ${\cal
F}=\bra{\psi}\rho\ket{\psi}$.  Choosing a particular path of links
${\cal P}$, and averaging this fidelity over all initial states, that
is all allowed values of $\alpha$ and $\beta$, gives $\bar{\cal
F}=(3+ \Pi_{i\in{\cal P}} \sqrt{4\phi_{0,i}\phi_{1,i}})/4$.  Because
each factor in the product can be associated with a link weight, this
problem is amenable to efficient algorithms.

Reference~\cite{DiF12} next generalises the problem slightly to the
case that the initial state of each link is one of a class of mixed
states that includes Werner states. In this case, it turns out that
each link must be assigned two independent link weights $x_i$ and
$y_i$ that depend on the parameters of the link state. Furthermore,
the average fidelity now has the form
\begin{equation}\label{twoprodmeasure}
 \bar{\cal F}= c + \prod_{i\in{\cal P}} x_i + \prod_{i\in{\cal P}} y_i.
\end{equation}
Consider adding a single node $Z$ that is connected
to the network only by one link to node $B$ and using
\eref{twoprodmeasure} to compute shortest paths.
It is easy to see that in general, the shortest path from $A$ to $Z$   does not include
the shortest path from $A$ to $B$. Thus, the problem no longer
possesses optimal substructure and cannot be solved by efficient algorithms
that rely on this property.

Similar questions arise in classical network engineering, and in
particular to their application to quantum networks. In general, one
might hope to find a measure of path length that one hand satisfies
the optimality principle, and on the other, whose minimisation is a
reasonable approximation of a more difficult global optimisation
problem. For instance, reference~\cite{VSLMN12} studies a model
optical network that creates and distributes Bell pairs via
simulations depending on many of parameters. The authors compare
optimisation based on various measures of the work done per link to
the global throughput. The simulations show that, within this model,
one can usually predict the path that simultaneously gives highest
throughput, and, by some measure of operations, the lowest work.


%% file: sections/conclusion.tex
\newpage
\setcounter{footnote}{0}
\section{Conclusion}
\markboth{\sc Conclusion}{}
\graphicspath{{conclusion/}}

In this review we have given an account of theoretical progress on the
distribution of entanglement in quantum networks. We have seen that
this inquiry has been driven by the experimental results on the
building blocks of entanglement distribution. These results are
impressive and promising, but also make evident fundamental and
technical barriers.  We have also seen that the task of distribution
is intimately connected with theoretical questions about the nature of
entanglement. In this setting, these questions are focused on the
extent to which it can be measured and inter-converted. While the
basic concepts (direct transmission, swapping, purification, error
correction) have been present for over twenty years, their application
to real networks is still in an exploratory phase.

The firsts steps to transmission of entanglement embodied in
a variety of physical systems have already been taken.
In fact, the most important application to
date, the distribution of quantum cryptographic keys, has been demonstrated.
Teleportation, as well, has been demonstrated in several
systems. But the problem of exponential decay of fidelity
has not been solved. Methods using the geometry of
higher-dimensional networks to effectively concentrate
entanglement from some sections for use in others shows
promise in overcoming this difficulty. At the same time,
they show strong connections with the theory of classical
networks and graphs, in particular with percolation theory.
Classical error correction also has
proven to be useful in this regard, with interesting quantum
connections, such as defining syndromes via weak
measurements that preserve information.

It is clear that the study of entanglement distribution on mixed-state
networks will be of prime importance. For instance, quantum
cryptography will continue to be one of the main
technologies driving research.  Quantum repeaters are
expected to become more robust, thereby
allowing small networks to form.  This will require new
protocols adapted to these small networks.  Eventually, we
expect to see application of pioneering work in entanglement
percolation and error correction. Finally, as the number of
nodes increases to the point that statistical methods can be
applied, we expect to see vigorous activity in the theory of
complex quantum networks.
